\documentclass[%
 reprint,
 amsmath,amssymb,
 aps,
prb,
floatfix,
]{revtex4-2}
\usepackage{graphicx}
\usepackage{dcolumn}
\usepackage{bm}
\usepackage[version=4]{mhchem}
\usepackage[caption=false]{subfig}
\usepackage{mathtools}
\usepackage{xcolor}

\usepackage{xr}

\usepackage[bitstream-charter]{mathdesign}
\usepackage[T1]{fontenc}

\begin{document}
\sloppy

\preprint{APS/123-QED}

\title{Finite-Temperature Ferroelectric Phase Transitions from Machine-Learned Force Fields}

\author{Kristoffer Eggestad}
\author{Ida C. Skogvoll}
\author{Øystein Gullbrekken}
\author{Benjamin A. D. Williamson}

\author{Sverre M. Selbach}%
 \email{selbach@ntnu.no}
\affiliation{Department of Materials Science and Engineering, Norwegian University of Science and Technology}%

\date{\today}

\begin{abstract}
 Simulating finite temperature phase transitions from first-principles is computationally challenging. Recently, molecular dynamics (MD) simulations using machine-learned force fields (MLFFs) have opened a new avenue for finite-temperature calculations with near-first-principles accuracy. Here we use MLFFs, generated using on-the-fly training, to investigate structural phase transitions in four of the most well-studied ferroelectric oxides; \ce{BaTiO3}, \ce{PbTiO3}, \ce{LiNbO3} and \ce{BiFeO3}. Only using the 0 K ground state structure as input for the training, the resulting MLFFs can qualitatively predict all the main structural phases and phase transitions, while the quantitative results are sensitive to the choice of exchange correlation functional with PBEsol found to be more robust than LDA and r$^2$SCAN. MD simulations also reproduce the experimentally observed order-disorder character of Ti displacements in \ce{BaTiO3}, the abrupt first order transitions of \ce{BiFeO3} and \ce{PbTiO3}, and the mixed order-disorder and displacive character of the ferroelectric transition in \ce{LiNbO3}. Finally, we discuss the potential and limitations of using MLFFs for simulating ferroelectric phase transitions. 

\end{abstract}

\maketitle

\section{Introduction}
\sloppy
Ferroelectric oxides, particularly perovskites, are of great technological and scientific interest and find applications in, e.g. sensors, actuators, transducers, and non-volatile memory technologies \cite{Wu2021,doi:10.1126/science.1129564,https://doi.org/10.1111/j.1151-2916.1999.tb01840.x}. The mechanisms of their ferroelectric phase transitions are imperative in order to understand the microscopic origins of spontaneous polarisation, and is thus enabling knowledge for designing new and improved ferroelectrics. Although first-principles computational approaches have been tremendously successful in elucidating fundamental properties and mechanisms at 0 K \cite{Cohen1992,PhysRevB.71.014113,VanAken2004,PhysRevB.83.094105}, the corresponding finite-temperature calculations have been prohibitively expensive. \textit{ab initio} molecular dynamics has been feasible only for relatively small systems where long-range electrostatic interactions, such as those in ferroelectrics, are not well captured or described.

 Machine-learned force fields (MLFFs), and in particular on-the-fly MLFFs \cite{on_the_fly_method,10.1063/5.0009491} have recently opened a new avenue for finite-temperature calculations with near-first-principles accuracy. Potentials are trained using a combination of molecular dynamics (MD), density functional theory (DFT), and machine learning to decide when to perform DFT steps and update force fields \cite{doi:10.1021/acs.jpclett.0c01061}. This approach can be utilised to train force fields for complex functional materials and allow for calculations of very large supercells that are not feasible using standard DFT. MLFFs have recently been used to predict phase transitions in a wide range of materials \cite{Verdi2021,PhysRevB.105.L060102,doi:10.1021/acs.jpcc.3c01542,PhysRevLett.122.225701,Liu2024,Klarbring2024,doi:10.1073/pnas.2410910122,doi:10.1021/acs.jpcc.3c03377,PhysRevB.108.235122,Fransson2023}, but ferroelectric materials and their phase transitions pose an additional challenge with their sensitivity to pressure and lattice parameters, and inherent balance between long- and short-range forces \cite{RevModPhys.77.1083,Bousquet2008}.
 
 Ferroelectric materials show an electric field-switchable spontaneous electric polarisation that emerges from a phase transition from a centrosymmetric high-temperature structure to a non-centrosymmetric low-temperature structure at the Curie temperature (T$_\text{C}$). \ce{BaTiO3}, \ce{PbTiO3}, \ce{BiFeO3}, and \ce{LiNbO3} are prototypical ferroelectric materials, with the first three belonging to the perovskite family, whereas \ce{LiNbO3} adopts a closely related structure. These four ferroelectric oxides span I-V, II-IV and III-III distributions of formal oxidation states, they include two examples of $d^0$-driven second-order Jahn-Teller mechanism of stabilising polarisation, and two examples of A-site 6$s^2$ lone pair-driven polarisation, the latter two of which we find one multiferroic with a non-$d^0$ B-site (\ce{BiFeO3}) and one with a $d^0$ B-site (\ce{PbTiO3}) cation. These four well-studied oxides thus constitute a robust testing ground for applying MLFFs to understand ferroelectric phase transitions and we briefly review the basics of their phase transitions below.

\ce{BaTiO3} undergoes three phase transitions upon heating; $R3m$ $\xleftrightarrow[]{\sim193\text{ K}}$ $Amm2$ $\xleftrightarrow[]{\sim 273\text{ K}}$ $P4mm$ $\xleftrightarrow[]{393\text{ K}}$ $Pm\Bar{3}m$ with the latter being the paraelectric aristotype phase \cite{PhysRev.76.1221}. All three phase transitions are first order and are associated with diverging dielectric constants and discontinuous volumes. While \ce{BaTiO3} is a prototypical displacive (soft-mode driven) ferroelectric, recent studies support a microscopic order-disorder picture rather than coherent, uniform displacements of \ce{Ti^{4+}} relative to the centre of mass of the \ce{TiO6} octahedra \cite{Comes:a07207,doi:10.1080/00150199808009173,doi:10.1080/00150199508221830,PhysRevLett.116.207602,doi:10.1080/00150199508221847,HULLER1969589}.

In contrast to \ce{BaTiO3}, \ce{PbTiO3}, \ce{LiNbO3} and \ce{BiFeO3} each exhibit only a single bulk ferroelectric phase. \ce{PbTiO3} undergoes a first-order transition, $P4mm$ $\xleftrightarrow[]{768\text{ K}}$ $Pm\Bar{3}m$\cite{T_C_PbTiO3}, which is accompanied by a large, discontinuous decrease in molar volume. \ce{LiNbO3} keeps its trigonal symmetry across the polar to non-polar ferroelectric transition, $R3c$ $\xleftrightarrow[]{\sim1480\text{ K}}$ $R\Bar{3}c$\cite{T_C_LiNbO3,T_C_LiNbO3_2}. 

The ferroelectric phase transition of \ce{BiFeO3} initially proved difficult to study experimentally \cite{https://doi.org/10.1002/adma.200802849,PhysRevB.77.014110} because of the high $T_C$, the volatility of Bi, and the chemical incompatibility towards common sample supporting materials like platinum, alumina (\ce{Al2O3}) and silica (\ce{SiO2})\cite{SELBACH20101205}. Arnold et al. conclusively established the nature of the ferroelectric transition of \ce{BiFeO3} to be $R3c$ $\xleftrightarrow[]{\sim1100\text{ K}}$ $Pbnm$\cite{arnold_BiFeO3}. The absence of a group-subgroup relationship between $Pbnm$ and $R3c$ implies that this transition must be first order, and this is corroborated by a large negative change in molar volume and discontinuous enthalpy and electrical resistivity across $T_C$\cite{arnold_BiFeO3,doi:10.1021/cm9021084}. An even more debated phase transition of \ce{BiFeO3} is that from paraelectric $Pbnm$ (the $\beta$-phase, $Pnma$ is the standard setting) to the elusive $\gamma$-phase, occurring just $\sim 10$ K below the peritectic decomposition temperature of $\sim 1200$ K \cite{PhysRevB.77.014110,https://doi.org/10.1002/adfm.201000118,SELBACH20101205}.

In this work, we generate and use machine-learned force fields (MLFFs) as interatomic potentials to accurately predict phase transitions in the well-known ferroelectric oxides \ce{BaTiO3}, \ce{PbTiO3}, \ce{LiNbO3} and \ce{BiFeO3}. We use only their ground state structure as input for training the force fields. Similar to 0 K first-principles calculations, we show that the quantitative results are strongly dependent on the choice of exchange-correlation functional. Finally, we discuss the potential and limitations of MLFFs for studying ferroelectrics.

\section{Methods}

 Density functional theory (DFT) and molecular dynamics (MD) calculations were performed using the VASP code \cite{VASP1,VASP2,VASP3}. Interactions between core and valence electrons (Ba: ($5s^2$, $5p^6$, $5d^{0.01}$, $6s^{1.99}$), Ti: ($3s^2$, $3p^6$, $3d^3$, $4s^1$), O: ($2s^2$, $2p^4$), Pb: ($5d^{10}$, $6s^2$, $6p^2$), Li: ($1s^2$, $2s^1$), Nb: ($4s^2$, $4p^6$, $4d^4$, $5s^1$), Bi: ($5d^{10}$, $6s^2$, $6p^3$), Fe: ($3s^2$, $3p^6$, $3d^7$, $4s^1$)) were described using the projector-argumented wave (PAW) method \cite{paw1,PAW}. A large plane-wave energy cutoff of 700 eV was used for all DFT calculations to avoid Pulay stresses associated with the large volume changes during the training of the MLFFs. Three sets of MLFFs were generated using LDA \cite{LDA}, PBEsol \cite{PBEsol} and r$^2$SCAN \cite{R2SCAN} and used for MD simulations to assess the consequences of generating interatomic potentials using different functionals. MD simulations were performed within the NpT ensemble using the Parrinello-Rahman barostat \cite{Parrinello_Rahman_1,Parrinello_Rahman_2} with a pressure of 1 atm and the Langevin thermostat \cite{sim_liq}. Langevin friction coefficients were set to 5 ps$^{-1}$ for all atoms. A fictitious mass of 1000 amu and a friction coefficient of 5 ps$^{-1}$ were used for the lattice degrees of freedom for all simulations. DFT calculations of \ce{BiFeO3} were initialised with a G-type antiferromagnetism, and a Hubbard U of 4 eV was used when generating MLFFs with LDA and PBEsol.

 An initial equilibration was first performed of the DFT optimised ground state structures using the on-the-fly scheme at temperatures of 50 K, 300 K, 500 K and 500 K for \ce{BaTiO3}, \ce{PbTiO3}, \ce{LiNbO3} and \ce{BiFeO3}, respectively. The equilibrations were performed using 50000 time steps with a step length of 1 fs. Supercells of 120 atoms, similar to a $2\times 2\times 1$ expansion of the conventional $R3c$ unit cell of \ce{LiNbO3} and \ce{BiFeO3}, were used for the initial equilibrations and all training. This supercell allows for anti-phase tilting of O octahedra in all directions and maximises the cell size without making the generation of interatomic potentials too computationally demanding. $\gamma$-centred k-point grids of $3\times 3\times 2$ were used for all DFT calculations. All interatomic potentials generated during the initial equilibrations were discarded.

 Generation of MLFFs was done by heating (100 steps per K) the equilibrated structures from the initial temperatures to temperatures well above the respective ferroelectric phase transition (\ce{BaTiO3}: 500 K, \ce{PbTiO3}: 1000 K, \ce{LiNbO3}: 1700 K and \ce{BiFeO3}: 1300 K) using the on-the-fly scheme. A cutoff radius of 8 Å was used to limit the atomic environments described by MLFFs. Generated MLFFs are uploaded to zenodo \cite{eggestad_2025_15784521}.

 Thereafter, production runs were done with equilibrium MD simulations at specific temperatures using the generated potentials. Production runs were carried out in series, meaning that simulations were performed from low to high temperature, where the final structure of the previous calculation was used as the initial structure of the next calculation. The length of all production runs was 40 ps with a time step of 1 fs. For all calculations, the lattice parameters converged within 20 ps, and only the final 20 ps of each simulation were analysed. For \ce{PbTiO3}, \ce{LiNbO3} and \ce{BiFeO3} a supercell containing 2000 atoms, similar to a $5\times 5 \times 4$ expansion of the conventional $Pbnm$ unit cell of \ce{BiFeO3}, was used for all production runs. Due to narrow temperature intervals of the different phases, simulations of \ce{BaTiO3} were performed using a 4900 atom supercell ($7\times 7 \times 5$ expansion) to reduce thermal fluctuations.

\section{Results}

 Lattice parameters as a function of temperature for \ce{BaTiO3}, \ce{PbTiO3}, \ce{LiNbO3} and \ce{BiFeO3} are displayed in Figure \ref{fig:BaTiO3_lattice_exp}, \ref{fig:PbTiO3_lattice_exp}, \ref{fig:LiNbO3_lattice_exp} and \ref{fig:BiFeO3_lattice_exp}, respectively. The top panels show average lattice parameters from equilibrium MD simulations using MLFFs generated with the PBEsol functional, while the bottom panels show experimental lattice parameters from the literature \cite{kay_lattice_BaTiO3,Shirane_PbTiO3,LiNbO3_lattice_para,arnold_BiFeO3}. In SI Figure S4, S7, S11 and S15 the experimental and simulated lattice parameters are plotted together as a function normalised to the T$_\text{C}$. Results from simulations using MLFFs generated with LDA and r$^2$SCAN are displayed in the SI Figure S6, S10, S14 and S18. 

\subsection{BaTiO$_3$}

 Equilibrium MD simulations, using MLFFs generated with PBEsol, qualitatively predict all the correct phases and transitions for \ce{BaTiO3} upon heating. Estimated phase transition temperatures are systematically about 70 K lower than experimentally observed transition temperatures \cite{kay_lattice_BaTiO3}. The calculated lattice parameters are in general about 0.01 Å larger than the experimental values, which can be attributed to the subtle underbinding tendency of PBEsol \cite{perdew_generalized_1996}. Except for the underestimated transition temperatures, the main difference between the experimental and our calculated lattice parameters is the change in volume over the phase transitions. Kay et al. \cite{kay_lattice_BaTiO3} reports a small increase in volume at the transition from $R3m$ to $Amm2$ and from $Amm2$ to $P4mm$. While our calculations predict a subtle decrease in volume at these transitions, the negative volume change across the ferroelectric transition is correctly reproduced by our simulations.

\begin{figure}[ht]
    \centering
    \hspace{-20pt}
    \includegraphics[width=0.9\linewidth]{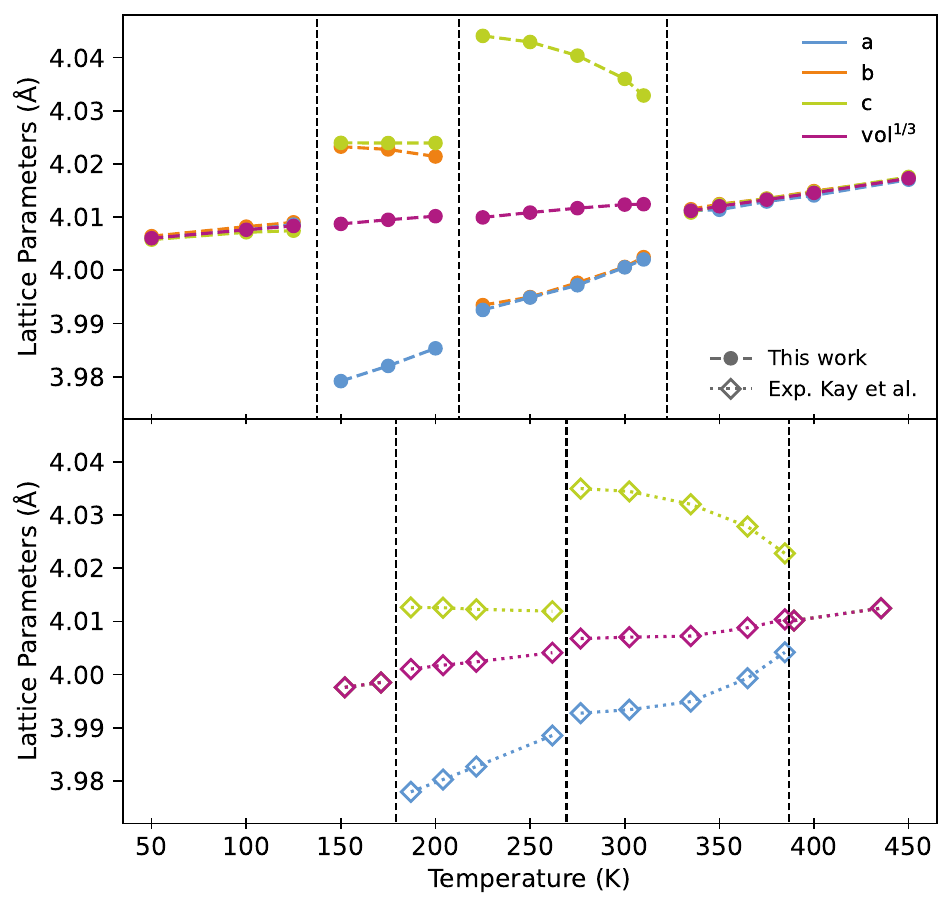}
    \caption{Evolution of lattice parameters as a function of temperature for \ce{BaTiO3}. The top panel displays predicted lattice parameters from MD simulations performed using MLFFs generated with the PBEsol functional, while the bottom panel shows experimental results from Kay et al. \cite{kay_lattice_BaTiO3}. The dashed and dotted vertical black lines indicate the phase transition temperatures.}
    \label{fig:BaTiO3_lattice_exp}
\end{figure}

 The LDA and r$^2$SCAN functionals also predict the correct phases and transitions. However, LDA underestimates the lattice parameters (SI Figure S6), due to its well-known tendency to overbind \cite{vandewalle_correcting_1999}, while r$^2$SCAN overestimates the lattice parameters. Furthermore, this is reflected in the estimated phase transition temperatures, where LDA predicts erroneously low transition temperatures, while transition temperatures from r$^2$SCAN are closer to the experimental values than those predicted from PBEsol (Figure \ref{fig:BaTiO3_lattice_exp}). 
 

\begin{figure}[ht]
    \centering
    \includegraphics[width=0.9\linewidth]{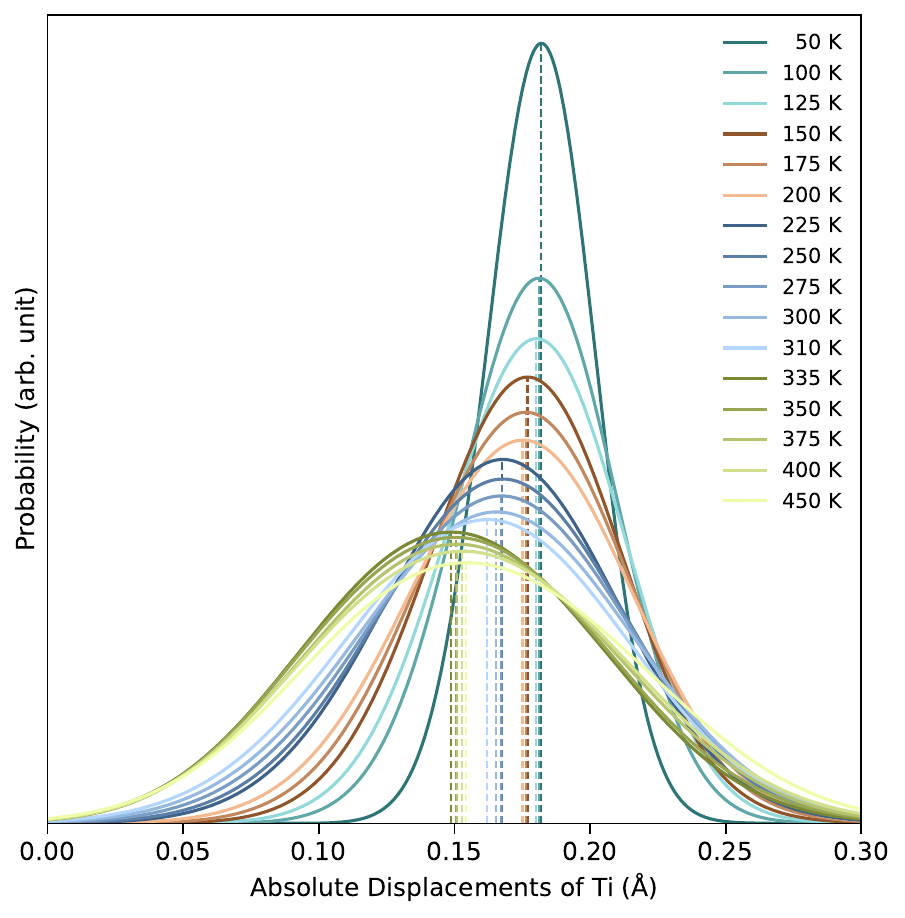}
    \caption{Probability distribution of absolute displacements of Ti in \ce{BaTiO3} from equilibrium MD simulations using MLFFs generated with PBEsol. The colours indicate the average space groups (turquoise: $R3m$, orange: $Amm2$, blue: $P4mm$, green: $Pm\Bar{3}m$) and the dashed vertical lines are guidelines for the peak positions.}
    \label{fig:abs_Ti_BaTiO3}
\end{figure}

 The probability distributions of the absolute Ti displacements in \ce{BaTiO3} at various temperatures, shown in Figure \ref{fig:abs_Ti_BaTiO3}, reveal a decrease in the Ti displacements upon heating up to the formation of the cubic phase. As the temperature increases further, the Ti displacements gradually increase. Each phase transition is marked by a step-like shift in the peak position, as indicated by the dashed vertical lines. Additionally, the probability distributions broaden with increasing temperature, reflecting enhanced thermal fluctuations.

\begin{figure*}[ht]
    \centering
    \includegraphics[width=0.9\linewidth]{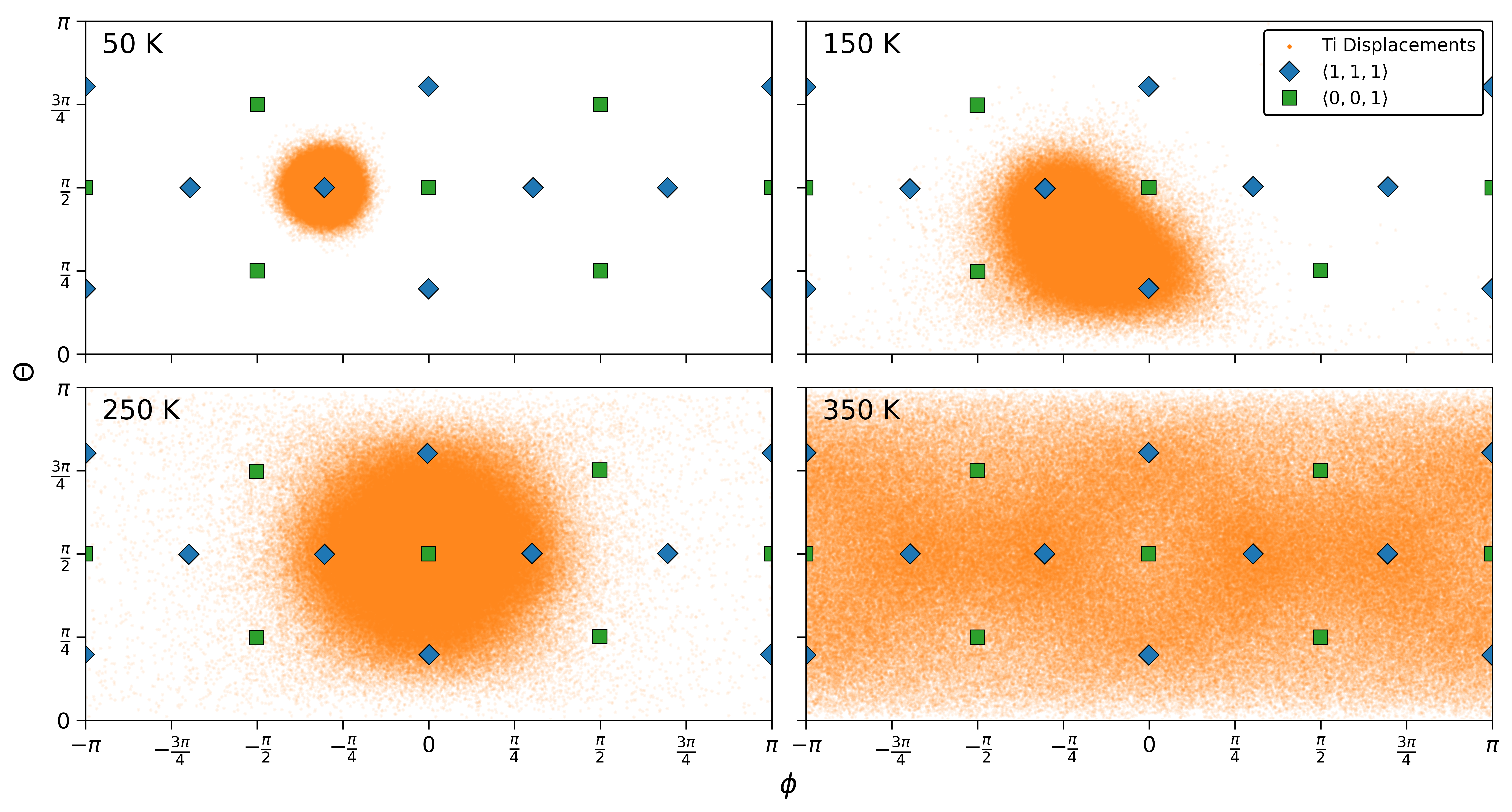}
    \caption{Mercator projections showing the direction of the Ti displacements in \ce{BaTiO3} at $50$ K, $150$ K, $250$ K and $350$ K extracted from equilibrium MD simulations using MLFFs generated with the PBEsol functional.  Small orange data points display the direction of the displacement of single Ti atoms from the centre of the respective O octahedra relative to the $\langle 1,0,0 \rangle$ (green squares) and $\langle 1,1,1 \rangle$ (blue diamonds) directions. Mercator projections for the orientation of the Ba displacements are shown in SI Figure S5.}
    \label{fig:mercator_Ti_BaTiO3}
\end{figure*}

 The direction of the Ti displacements at 50 K, 150 K, 250 K and 350 K are displayed using Mercator projections shown in Figure \ref{fig:mercator_Ti_BaTiO3}. At 50 K, all Ti atoms are displayed in one of the $\langle 1,1,1 \rangle$ directions, consistent with the $R3m$ space group. At 150 K, Ti atoms are displaced in two different $\langle 1,1,1 \rangle$ directions, and the average orientation of the displacements gives rise to the direction of the macroscopic polarisation observed in orthorhombic \ce{BaTiO3}. Moreover, at 250 K, Ti atoms are disordered in between four different $\langle 1,1,1 \rangle$ directions, where the average gives rise to tetragonal \ce{BaTiO3}. At 350 K, the orientation of the Ti displacements is disordered between all $\langle 1,1,1 \rangle$ directions. A similar cation off-centring has, for instance, been reported by K. Page et al. for Nb-substituted cubic \ce{BaTiO3} \cite{PhysRevLett.101.205502}. Thermal energy gives broader distributions, making the disorder between $\langle 1,1,1 \rangle$ directions at higher temperatures less obvious. 

 The temperature-dependent evolution of Ti displacements, as visualised in the Mercator projections, clearly supports the the order-disorder picture of phase transitions in \ce{BaTiO3} \cite{first_disorder, COMES1968715, Comes:a07207, doi:10.1080/00150199808009173, doi:10.1080/00150199508221830, PhysRevLett.116.207602}. Similar to work by Holma et al. \cite{doi:10.1080/00150199508221847}, our calculations do not support the formation of static linear chains of correlated displacements proposed by Comes et al. \cite{COMES1968715,Comes:a07207}, but are instead more consistent with the dynamic description by Hüller et al. and Harada et al. \cite{HULLER1969589,Hüller1969,PhysRevB.4.155}.

\subsection{PbTiO$_3$}

\begin{figure}[ht]
    \centering
    \hspace{-20pt}
    \includegraphics[width=0.9\linewidth]{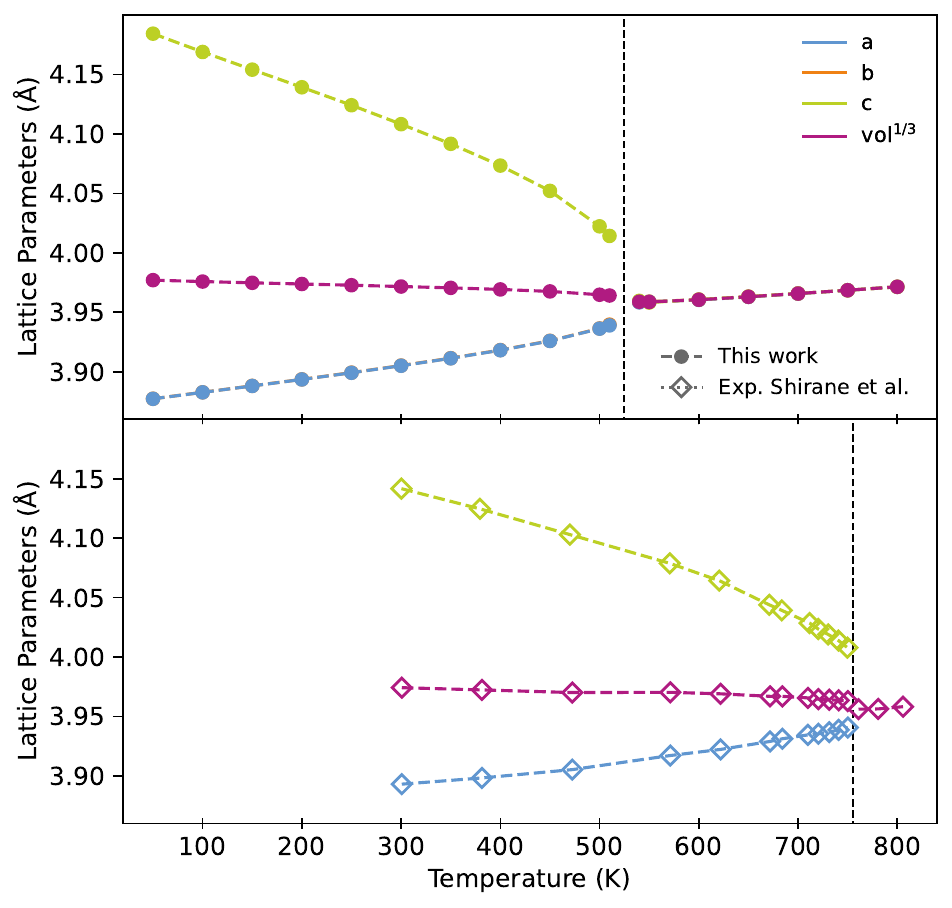}
    \caption{Evolution of lattice parameters as a function of temperature for \ce{PbTiO3}. The top panel displays predicted lattice parameters from the MD simulations using MLFFs generated with the PBEsol functional, while the bottom panel shows experimental results from Shirane et al. \cite{Shirane_PbTiO3}. The dashed and dotted vertical black lines indicate the phase transition temperatures.}
    \label{fig:PbTiO3_lattice_exp}
\end{figure}

 Simulations of \ce{PbTiO3} using PBEsol, Figure \ref{fig:PbTiO3_lattice_exp}, show the same lattice parameter evolution as reported from the seminal experiments by Shirane et al. \cite{Shirane_PbTiO3}, albeit a significantly lower transition temperature of about 230 K. Both simulations and experiments show a slowly decreasing cell volume with increasing temperature up until the phase transition followed by a significant reduction at the transition temperature. Thereafter, a slow increase in volume with increasing temperature is observed. Similar results are shown by LDA and r$^2$SCAN and are displayed in SI Figure S10. LDA, again, severely underestimates the lattice parameters and the phase transition temperature, while r$^2$SCAN overestimates the lattice parameters.


 Probability distributions of the displacements of Pb and Ti from the centre of oxygen polyhedra, in SI Figure S1, show that both cations are significantly displaced at all temperatures. These distributions show a clear change in trend and jump in peak position from 500 K to 550 K (see also plotted peak positions in SI Figure S2). Similar to the ferroelectric transition in \ce{BaTiO3}, at the T$_\text{C}$, the average atomic displacements from polyhedral centres reverts from decreasing, go through a minimum and start increasing upon further heating.


 Contrary to \ce{BaTiO3}, in the $P4mm$ phase, cation displacements in \ce{PbTiO3} are not disordered and only displaced along one $\langle 0,0,1 \rangle$ direction, as reported experimentally by Kwei et al. \cite{doi:10.1080/00150199508221830} and supported by the Mercator projections in SI Figure S8 and S9. Above the T$_\text{C}$, in the paraelectric phase, the displacements are fully disordered in all directions.


 The average B-cation displacement was also fitted as a function of the temperature to the power law $A(T_C -T)^\beta$, giving a value of 0.27 for the critical exponent $\beta$. This is reasonably close to the theoretical prediction of 0.25 for a tricritical phase transition, as expected for \ce{PbTiO3}. For more details on the calculation of critical exponents, see SI Section II.

\subsection{LiNbO$_3$}

\begin{figure}[ht]
    \centering
    \hspace{-20pt}
    \includegraphics[width=0.9\linewidth]{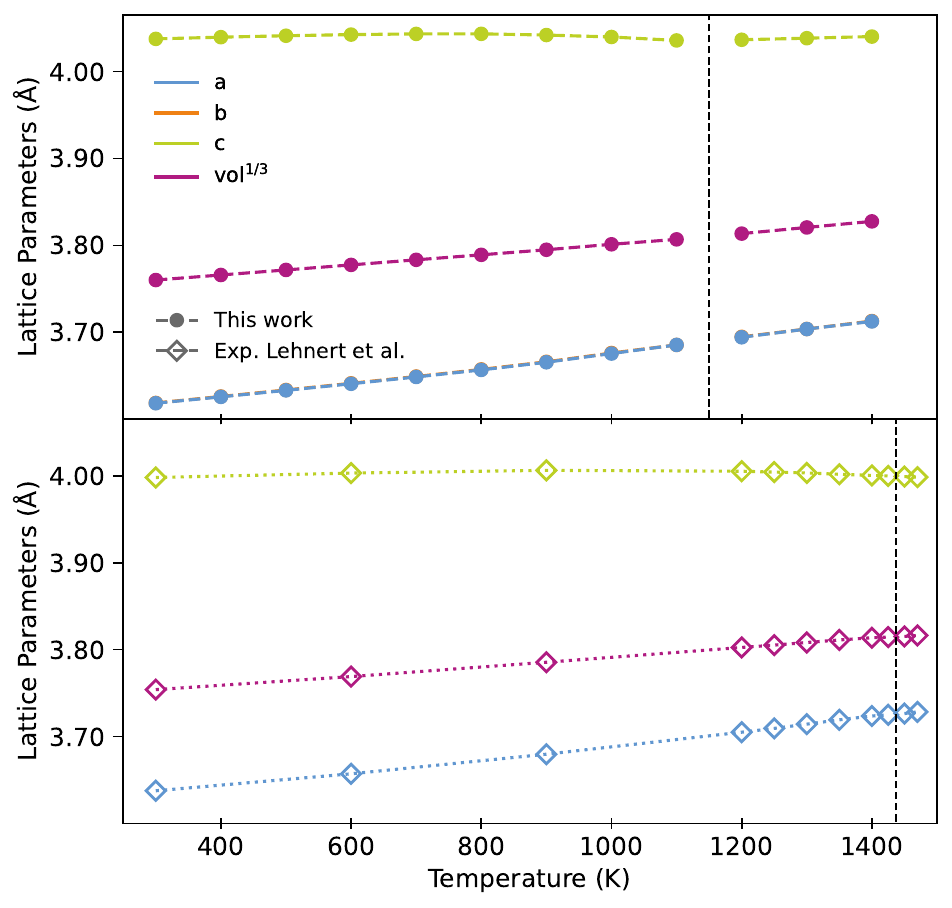}
    \caption{Evolution of lattice parameters as a function of temperature for \ce{LiNbO3}. Lattice parameters are converted from the hexagonal conventional unit cell to a pseudocubic cell by dividing a and c by $\sqrt{2}$ and $\sqrt{12}$, respectively. The top panel displays predicted lattice parameters from the MD simulations using MLFFs generated with the PBEsol functional, while the bottom panel shows experimental results from Lehnert et al. \cite{LiNbO3_lattice_para}. The dashed and dotted vertical black lines indicate the phase transition temperatures.}
    \label{fig:LiNbO3_lattice_exp}
\end{figure}

 Similar to lattice parameters from experiments by Lehnert et al. \cite{LiNbO3_lattice_para} and Sugii et al. \cite{SUGII1976199}, the calculated \textit{a} lattice parameter increases linearly up to and beyond the phase transition, while the \textit{c} parameter starts decreasing when approaching the phase transition temperature (Figure \ref{fig:LiNbO3_lattice_exp}). Due to the second-order nature of the phase transition in \ce{LiNbO3}, the Curie temperature is very difficult to estimate exclusively from the changes in lattice parameters alone. However, similar to the probability distributions for cation displacements in \ce{BaTiO3}, \ce{PbTiO3} and \ce{BiFeO3} (SI Figure S1), there is a clear abrupt change in the evolution of the absolute displacements of Li and Nb from 1100 K to 1200 K, indicating the T$_\text{C}$ to be in between these temperatures. Using MLFFs generated with PBEsol, the predicted T$_\text{C}$ is about 300 K lower than the experimentally observed T$_\text{C}$. With LDA and r$^2$SCAN, lattice parameters are again underestimated and overestimated, respectively, see SI Figure S14.

\begin{figure*}[ht]
    \centering
    \subfloat[]{\includegraphics[width=0.45\linewidth]{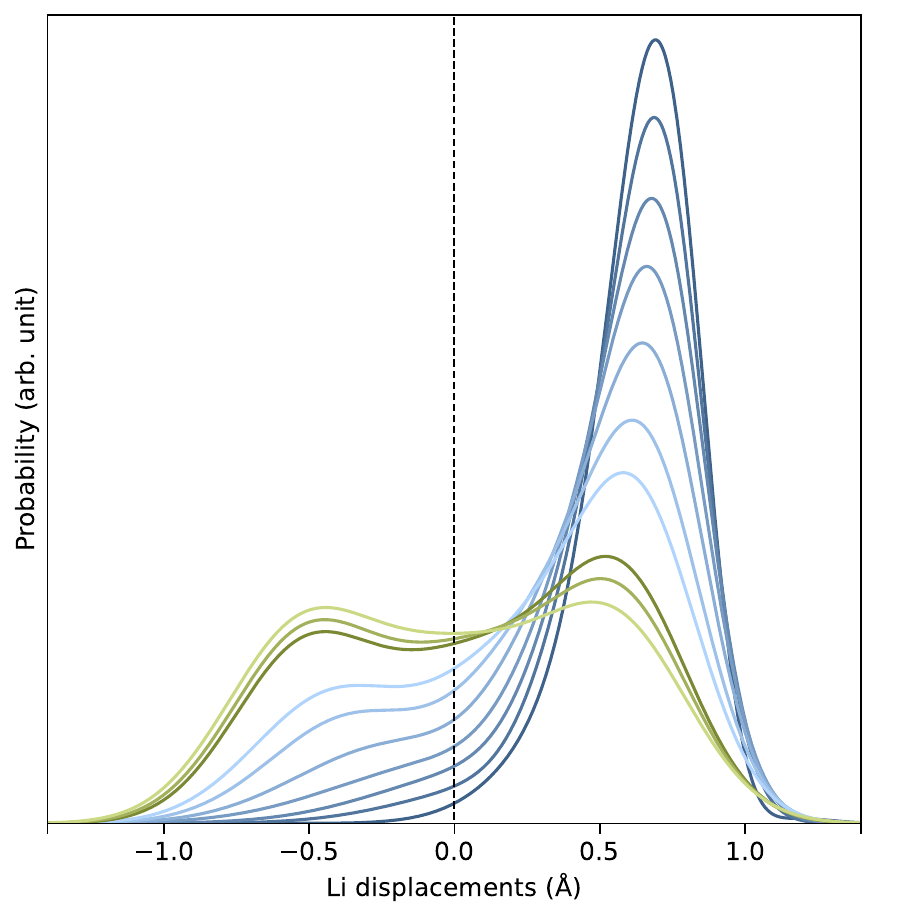}}
    \quad\quad\quad\quad
    \subfloat[]{\includegraphics[width=0.45\linewidth]{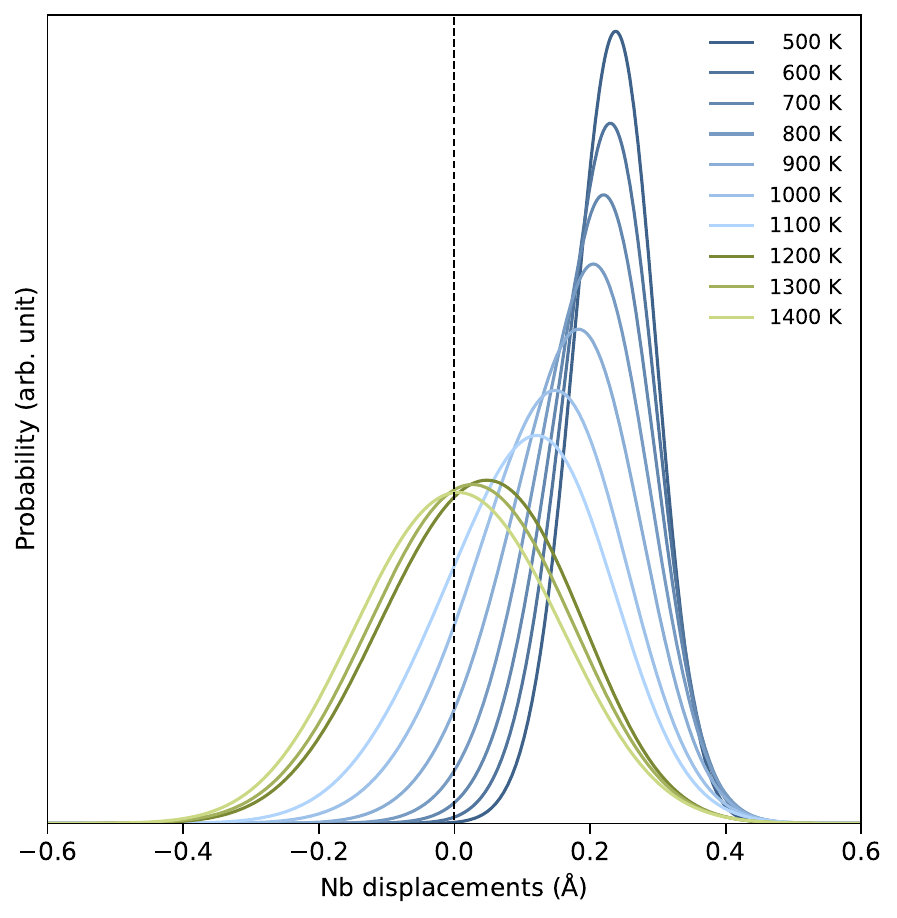}}
    \caption{Probability distributions, from equilibrium MD simulations using MLFFs generated with the PBEsol functional, of Li (a) and Nb (b) displacements in the polar direction [1,1,1] direction. The colours indicate the space groups (blue: $R3c$, green: $R\Bar{3}c$) and the gradient shows the temperature.}
    \label{fig:LiNbO3_displacements}
\end{figure*}


 Below the T$_\text{C}$, both Li and Nb are displaced along the same $\langle 1,1,1 \rangle$ direction. Displacements along this [1,1,1] direction, displayed in Figure \ref{fig:LiNbO3_displacements}, highlight the differences between Li and Nb when approaching the phase transition temperature. Li, Figure \ref{fig:LiNbO3_displacements} (a), clearly show an order-disorder transition, while Nb show a displacive transition. Similar distributions and results have been reported from \textit{ab initio} MD by Sanna et al. \cite{6306010}. The order-disorder and the displacive transitions are also clearly shown in the Mercator projections in SI Figure S12 and S13, respectively. Mixed displacive and order-disorder characteristics of a ferroelectric transition have also been observed for other uniaxial ferroelectrics like hexagonal manganites, \ce{h-RMnO3} \cite{PhysRevX.9.031001}.
 
 From 1100 K to 1200 K, at the estimated T$_\text{C}$, there is a jump in the evolution of the displacements of both Li and Nb. However, the distributions of Li and Nb indicate that the material is still weakly polar, which could be explained by insufficient thermal energy to reach full ergodicity within the limited simulation time analysed.

 A power law fitting of the average B-cation displacement resulted in a $\beta$-value of 0.34, in very good agreement with the expected value for a second-order ferroelectric phase transition, as well as the experimental value of $\beta=0.355$ measured for isostructural \ce{LiTaO3} \cite{Hushur2007}.

\subsection{BiFeO$_3$}

\begin{figure}[ht]
    \centering
    \hspace{-20pt}
    \includegraphics[width=0.9\linewidth]{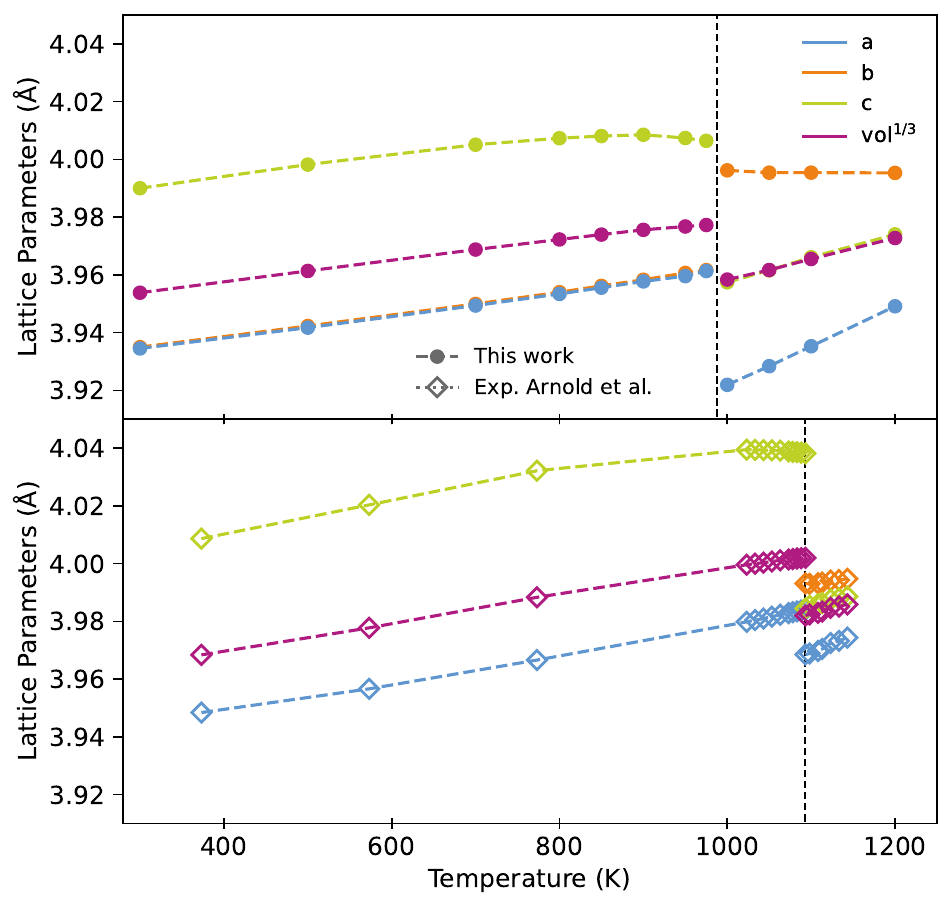}
    \caption{Evolution of lattice parameters as a function of temperature for \ce{BiFeO3}. Similar to \ce{LiNbO3}, in the rhombohedral phase, lattice parameters are converted from the hexagonal conventional unit cell to a pseudocubic cell by dividing a and c by $\sqrt{2}$ and $\sqrt{12}$, respectively. Lattice parameters from the orthorhombic structure are converted by dividing a/b and c by $\sqrt{2}$ and $2$, respectively. The top panel displays predicted lattice parameters from the MD simulations in this work, while the bottom panel shows experimental results from Arnold et al. \cite{arnold_BiFeO3}. The dashed and dotted vertical black lines indicate the phase transition temperatures.}
    \label{fig:BiFeO3_lattice_exp}
\end{figure}

 Both the non-linearity and the thermal stability of the ferroelectric $R3c$ phase of \ce{BiFeO3} is fairly well described by the PBEsol-based simulations, including the premonitory phenomenon of the lattice parameter c, parallel to the spontaneous polarisation (similar to \ce{LiNbO3}), starting to contract as T$_\text{C}$ is approached \cite{arnold_BiFeO3}. The large negative and discontinuous volume change across the ferroelectric phase transition of \ce{BiFeO3} is also reproduced, as shown in Figure \ref{fig:BiFeO3_lattice_exp}, in agreement with Arnold $et al.$ \cite{arnold_BiFeO3}. 

 At the T$_\text{C}$, MLFFs generated using LDA and PBEsol predict a transition from rhombohedral $R3c$ to orthorhombic $Pbnm$, with LDA again underestimating the lattice parameters. r$^2$Scan, on the other hand, gives a qualitatively different result and shows a transition from $R3c$ directly to cubic $Pm\Bar{3}m$. Using LDA and PBEsol, the difference between the calculated \textit{a}, \textit{b}, and \textit{c} lattice parameters in the $Pbnm$ phase is substantially larger than experimentally reported. However, DFT-optimised lattice parameters at 0 K give similar values, see SI Table S1.


 The volume contraction of the unit cell is also reflected in the distributions of absolute cation displacements, shown in SI Figure S1, with drastic changes to the peak positions for both the Bi and Fe distributions. Except for the abnormal changes at the T$_\text{C}$, these peak positions show only a weak temperature dependence (SI Figure S2). These distributions show increased broadening as a function of increasing temperature for the displacements of both Bi and Fe.


 The Mercator projections for Bi and Fe, in SI Figure S16 and S17, are relatively similar to the Mercator projections for \ce{LiNbO3}, with the main differences being a very abrupt phase transition in \ce{BiFeO3} and the fact that the displacements of Bi are not disordered, but instead ordered in an alternating fashion in the high-temperature phase, in agreement with the structural degrees of freedom of the paraelectric $Pbnm$ phase.

\section{Discussions}

 The observed underestimation of phase transition temperatures may be caused by several factors. Finite size effects, due to the use of limited cell sizes and periodic boundary conditions during both generation of MLFFs and production runs, will prevent the system from behaving as it would have done at the thermodynamic limit. As ferroelectric phase transitions are governed by long-range collective displacements of atoms, local fluctuations of the positions of individual atoms may easily destabilise the ferroic order. Moreover, MLFFs in VASP only describe the environment around atoms closer than a specific cutoff radius (default: 8 Å), and thus, long-range interactions are not directly described by the generated MLFFs. Furthermore, ferroelectric materials often experience decreasing and more diffuse T$_\text{C}$ with decreasing particle sizes \cite{PhysRevB.37.5852,Ishikawa_1996,doi:10.1021/nl052538e,doi:10.1126/science.1098252,ANNAPUREDDY2012206,doi:10.1021/ja0758436,PhysRevB.52.13177,PhysRevLett.105.185501}. Strain from the surfaces hinders the collective displacements and suppresses the ferroelectric phase. 

 A larger cutoff radius as well as larger supercells for both production runs and training of the interatomic potentials may mitigate the present errors in transition temperatures. However, increasing the cell size significantly will be computationally expensive. Supporting this notion, it has been reported that for \ce{LiNbO3} performing \textit{ab initio} MD with a 80 atom supercell underestimates the T$_\text{C}$ significantly \cite{6306010}, while a 640 atom supercell gives a T$_\text{C}$ very close to the experimental value \cite{sanna2}. Additionally, generating MLFFs using different exchange correlation functionals has proven to give vastly different transition temperatures \cite{PhysRevB.105.L060102,doi:10.1021/acs.jpcc.3c01542}, and thus, finding the optimal functional for the system is crucial. As of now, there is no "safe" choice of a single functional for training MLFFs for predicting ferroelectric phase transitions.

 In \ce{BaTiO3}, the spontaneous polarisation is solely due to second-order Jahn-Teller distortions caused by partial covalency between Ti 3$d$ and O 2$p$. This results in Ti being displaced towards one of the faces of the O octahedron, causing polarisation in one of the $\langle 1,1,1 \rangle$ directions. The other directions of polarisation observed experimentally are just combinations of polarisation in multiple $\langle 1,1,1 \rangle$ directions, as visualised with the Mercator projections in Figure \ref{fig:mercator_Ti_BaTiO3}. \ce{PbTiO3} also has Ti $d^0$, but only shows polarisation in the $\langle 0,0,1 \rangle$ directions, reflecting that the lone pair, caused by hybridisation of Pb 6$s$ and 6$p$ states, dictates the direction of the polarisation and the strongly coupled unit cell distortion. Furthermore, G. Laurita et al. showed that the introduction of lone pair cations, by Sn-substitution in \ce{BaTiO3}, drives local ordering and results in a behaviour closely resembling that of \ce{PbTiO3} \cite{PhysRevB.92.214109}.


 Similar to \ce{BaTiO3}, polarisation in \ce{LiNbO3} is only caused by d$^0$ second-order Jahn-Teller distortions. In spite of this, \ce{LiNbO3} only show two different phases ($R3c$ and $R\Bar{3}c$). The small size of Li atoms stabilises antiferrodistortive rotations and thus also the $R3c$ and $R\Bar{3}c$ phases. This also explains the order-disorder character of Li observed in Figure \ref{fig:LiNbO3_displacements} (a) and in the Mercator projections in SI Figure S12. Li is too small to be stable in the centre of the O polyhedra even at temperatures above the T$_\text{C}$.

 The simulated lattice parameters differ quantitatively from the experimental for the high-temperature $Pbnm$ phase of \ce{BiFeO3}, with the simulated unit cell being more metrically distorted. This is one possible reason why a transition to the experimentally observed $\gamma$-phase is not found in our simulations. Another obvious reason for the discrepancy is the possible evaporation of Bi and O at high temperatures, which would be very challenging to accurately predict and represent computationally. Our current simulations within the isothermal-isobaric NpT ensemble do not allow exchange of particles, and extending the MLFF approach to the grand canonical ensemble implies several challenges we will not discuss further here. Experiments where 30 molar \% of Fe has been replaced by Mn to lower the structural phase transition temperatures agree on the $T_C$ and the symmetries of the ferroelectric and the paraelectric phases, but not on the $\gamma$-phase \cite{PhysRevB.79.214113,PhysRevB.87.224109}, emphasizing the challenging nature of this particular phase transition and the need for further work to clarify e.g. the possible role of non-stoichiometry.

 In general, except for special situations, the choice of functional does not change the results qualitatively. Moreover, disregarding thermal expansion, the generated MLFFs closely follow lattice parameters from regular DFT calculations with the respective functional. LDA is known to underestimate lattice parameters \cite{PhysRevB.79.085104}, which is reflected in the MD simulations using MLFFs generated with LDA. This leads to further underestimation of phase transition temperatures, which can make it difficult to separate and study phases that experimentally show narrow temperature windows such as \ce{BaTiO3}. Similarly, r$^2$SCAN has been shown to, on average, overestimate lattice parameters \cite{R2SCAN}, and our MD simulations, performed using MLFFs generated with r$^2$SCAN, also tend to overestimate lattice parameters. Furthermore, these MLFFs fail to correctly describe both phase transitions in \ce{BiFeO3}, which may partly be due to the overestimated lattice parameters. Our results suggest that functionals that agree well with experimental values at the DFT level also result in MLFFs that accurately describe lattice parameters at elevated temperatures. Therefore, PBEsol is, in most situations, a good choice in terms of both speed and accuracy. 
 
 A major advantage of using MD simulations is access to massive supercells. This can allow for calculations of larger phenomena such as surfaces, grain boundaries, superstructures, amorphous materials, nanoparticles, and domain walls and their interactions with defects. The flexibility of using on-the-fly machine-learned interatomic potentials makes it easy to apply the methodology shown in this work to phenomena such as those mentioned above, as well as to completely new and unknown systems. The balance between long-range and short-range electrostatic forces, and the dipole-stabilizing second-order Jahn-Teller effect (or improper ferroelectric mechanisms) is particularly challenging for MLFFs, necessarily leading to difficult trade-offs between cost and accuracy. However, this also implies that ferroelectrics are ideal test beds for further method development of MLFF approaches to achieve near-first-principles accuracy at a fraction of the cost.  

\section{Conclusions}

 We have demonstrated that we can accurately predict lattice parameters and phase transitions of ferroelectric materials using MLFFs trained from first-principles calculations. Qualitatively, our calculations get all the correct phases and transitions, except for the $\beta$-$\gamma$ transition in \ce{BiFeO3}, although predicted phase-transition temperatures are in general significantly lower than what is observed experimentally. This method can be used to resolve and describe more subtle structural changes, such as the magnitude of the B-cation displacements and the orientation of these displacements.

\begin{acknowledgments}
 The simulations were performed with resources provided by Sigma2 - the National Infrastructure for High-Performance Computing and Data Storage in Norway through project NN9264K. Support for this project was provided from the Research Council of Norway through projects 302506, 301954, and 354614.
\end{acknowledgments}

\bibliographystyle{unsrt}
\bibliography{references}

\end{document}


\preprint{APS/123-QED}

\title{SI - Finite-Temperature Ferroelectric Phase Transitions from Machine-Learned Force Fields}

\author{Kristoffer Eggestad}

\author{Ida C. Skogvoll}
\author{Øystein Gullbrekken}
\author{Benjamin A. D. Williamson}

\author{Sverre M. Selbach}%
 \email{selbach@ntnu.no}
\affiliation{Department of Materials Science and Engineering, Norwegian University of Science and Technology}%

\date{\today}

\maketitle
\beginsupplement

All data presented in figures in the SI, unless explicitly stated, show results from MD simulations using MLFFs generated with the PBEsol functional. In general, the different functionals lead to qualitatively similar results, and thus, presenting all the data from the three different functionals investigated in this work is somewhat superfluous. We chose PBEsol to represent our results as it has been shown to give lattice parameters close to experimental values \cite{AMOUNAS2023415002,D4TC02856B,doi:10.1021/acs.jpcc.6b08548}. The main differences between the different functionals are the calculated lattice parameters and are highlighted in SI Figure \ref{fig:BaTiO3_funcs}, \ref{fig:PbTiO3_funcs}, \ref{fig:LiNbO3_funcs} and \ref{fig:BiFeO3_funcs}.

\section{A- and B-cation Displacements}

\begin{figure}[ht]
    \centering
    \includegraphics[width=0.9\linewidth]{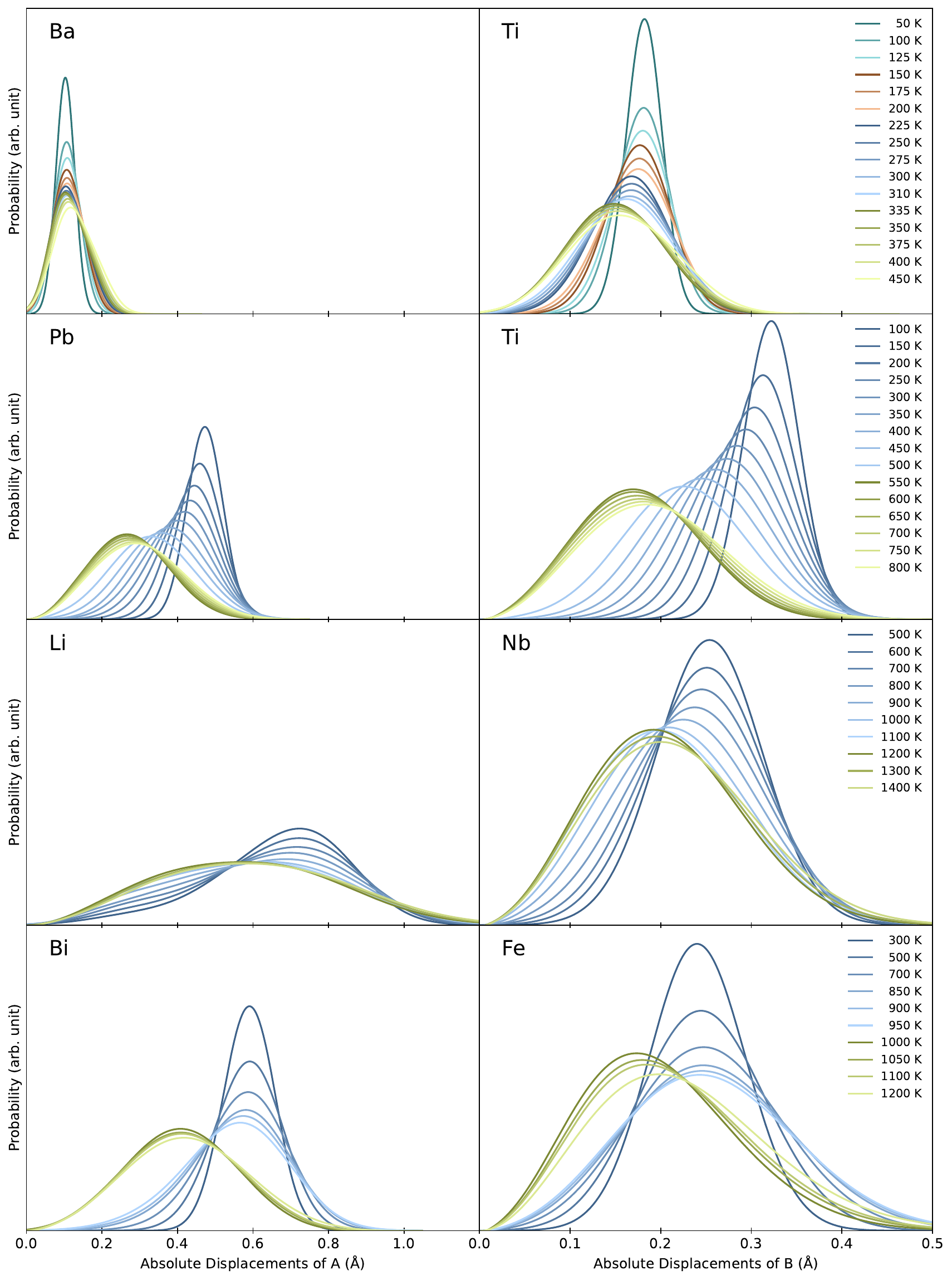}
    \caption{Probability distributions of the absolute displacements of A- and B-cations from O polyhedral centres for \ce{BaTiO3}, \ce{PbTiO3}, \ce{LiNbO3} and \ce{BiFeO3}. The colours indicate different phases.}
    \label{fig:abs_all}
\end{figure}

\begin{figure}[ht]
    \centering
    \includegraphics[width=0.65\linewidth]{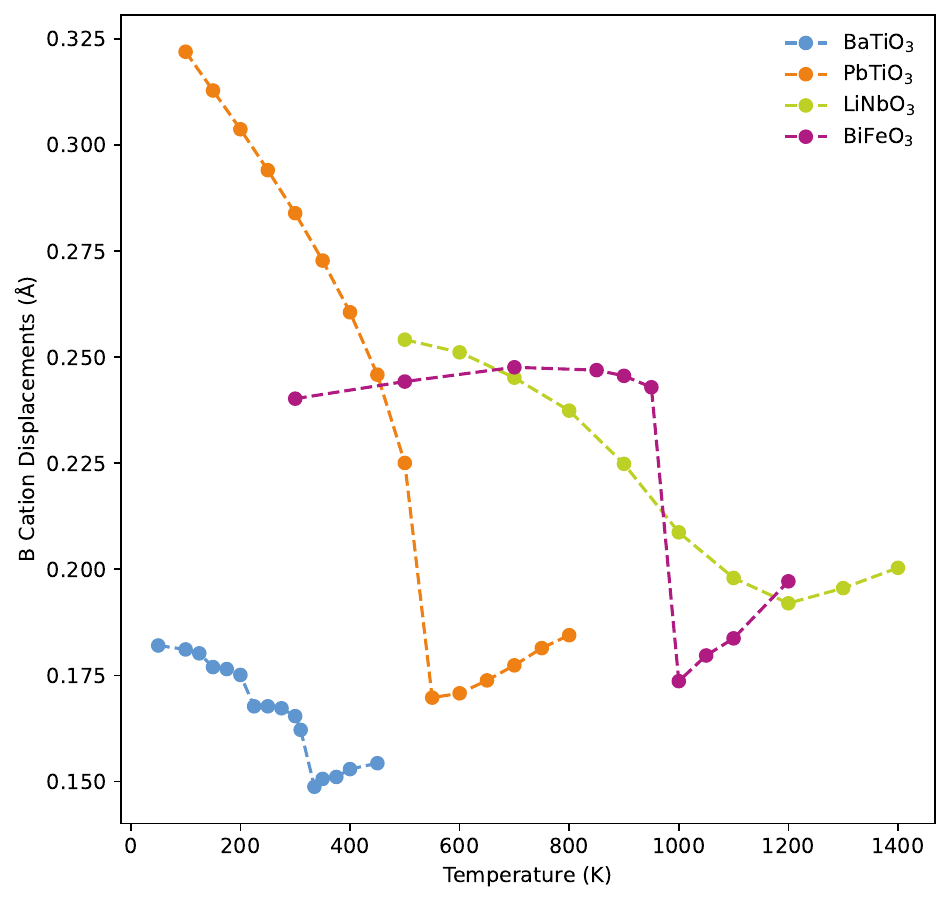}
    \caption{Peak position of distributions absolute B-cation displacements from the O octahedral centres seen in SI Figure \ref{fig:abs_all}.}
    \label{fig:peak_B}
\end{figure}

Figure \ref{fig:abs_all} shows probability distributions of the absolute displacements of cations from the O polyhedra centre calculated using the MLFFs generated using the PBEsol functional. From top to bottom, the panels show distributions for \ce{BaTiO3}, \ce{PbTiO3}, \ce{LiNbO3} and \ce{BiFeO3}. The left and right panels display A-and B-cation displacements, respectively. The different colours indicate different phases and the brightness shows the temperature from which the distribution is extracted. The brightness of the colours increases with increasing temperature within the same phase. The peak position of the B-cation distributions is shown in SI Figure \ref{fig:peak_B}.

\ce{PbTiO3} shows the largest B-cation displacements in the ferroelectric phase. In the paraelectric phase, the magnitude of the displacements is closely linked to the temperature, consequently \ce{BaTiO3} shows the smallest displacements while \ce{LiNbO3} shows the largest displacements. All compounds in the ferroelectric phase, except \ce{BiFeO3}, show a clear reduction in B-cation displacements with increasing temperature. In contrast, the Fe displacements in \ce{BiFeO3} increase slightly at low temperatures and start decreasing when approaching the phase transition temperature. In the paraelectric phase, displacements increase approximately linearly with increasing temperature for the investigated materials.

\clearpage

\section{Critical exponents}
\label{critical}
\begin{figure}[ht]
    \centering
    \includegraphics[width=0.65\linewidth]{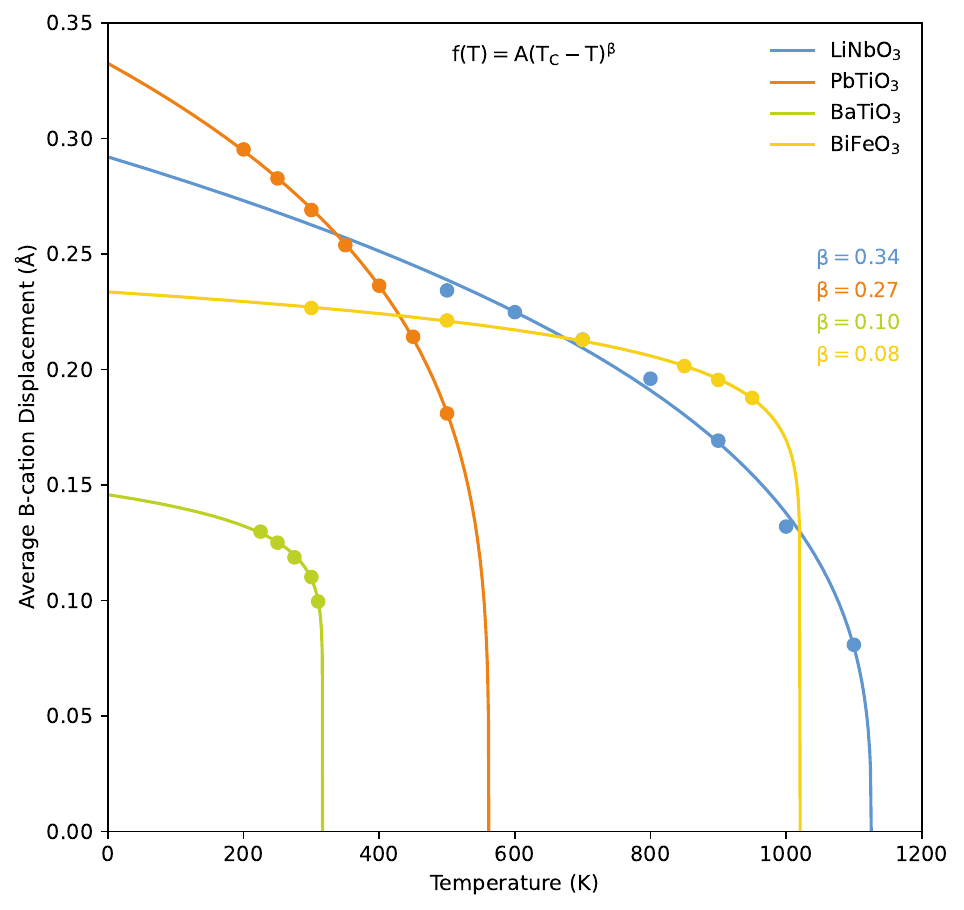}
    \caption{Average B-cation displacements as a function of temperature up to the ferroelectric phase transitions together with curve fits used to determine the critical exponents.}
    \label{fig:critical_exp}
\end{figure}

To verify the correct characteristics of the ferroelectric to paraelectric phase transitions, we also investigated the power law behaviour of the order parameters and determined the critical exponent $\beta$. This value can indicate whether the transition is strictly first-order, second-order or shares features of both. For each material, the average B-cation displacement as a function of the temperature was fitted to the expression $f(T) = A(T_C -T)^\beta$. For the $\beta$-values and the curve fit, see Fig. \ref{fig:critical_exp}. For \ce{BaTiO3} and \ce{BiFeO3}, the values are very small, consistent with an abrupt jump in the order parameter, as expected from first-order transitions. This also means that the values do not correspond to any theoretical models. 

However, for \ce{PbTiO3} and \ce{LiNbO3} the values fit well with experiment and theoretical predictions. For \ce{PbTiO3}, $\beta=0.27$ is close to the theoretical value of 0.25 for a tricritical phase transition, which represents the boundary between first-order and second-order. This agrees with the first-order transition of \ce{PbTiO3} becoming second-order under applied pressure \cite{sani2002pressure, rossetti2005specific}. A $\beta$-value of 0.34 is the observed and theoretically predicted value for a second-order ferroelectric transition, and has been measured to be 0.355 for isostructural \ce{LiTaO3} \cite{Hushur2007}. 

Fitting the strain instead of the polarization, in the form of the tetragonal distortion $(c-a)/c$, resulted in essentially the same values. Fitting to other critical exponents than $\beta$ was not attempted due to the difficulty of fitting a diverging parameter so close to the critical point. 

\clearpage

\section{B\lowercase{a}T\lowercase{i}O$_{3}$}

\begin{figure}[ht]
    \centering
    \includegraphics[width=0.65\linewidth]{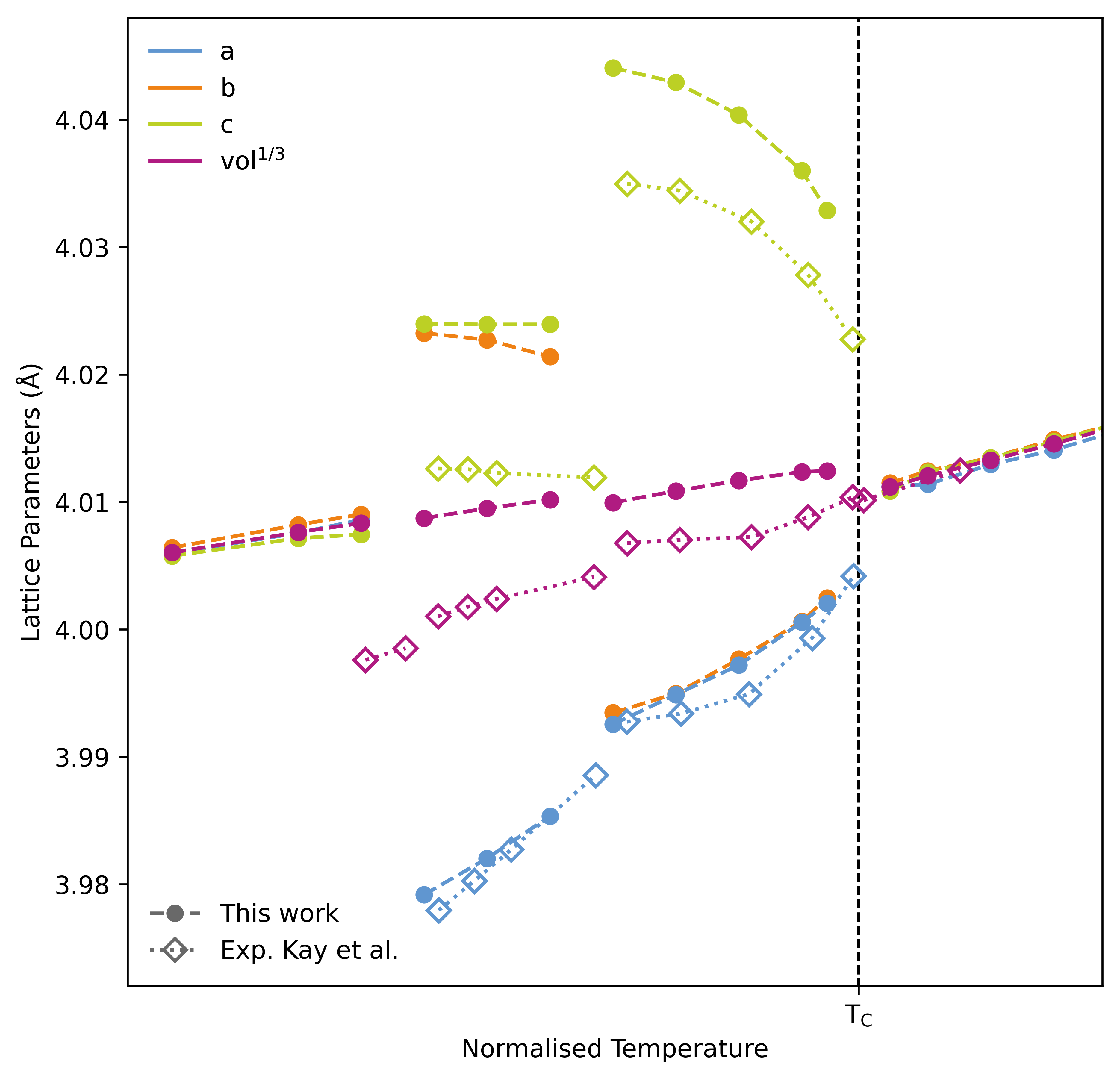}
    \caption{Lattice parameters, calculated using PBEsol, compared to experimental lattice parameters by Kay et al. \cite{kay_lattice_BaTiO3} as a function of temperature normalised to T$_\text{C}$.}
    \label{fig:lattice_BaTiO3_comb}
\end{figure}

SI Figure \ref{fig:lattice_BaTiO3_comb} shows lattice parameters, calculated using PBEsol, plotted on top of experimental lattice parameters as a function of temperature normalised to T$_{\text{C}}$. The trend of the lattice parameters is relatively similar for all phases. For the $R3m$ phase the calculated lattice parameters are about 0.01 Å longer. The $Amm2$ and $P4mm$ phases show good overlap for the a-parameter, while our calculations tend to overestimate the longer lattice parameters. For the $Pm\Bar{3}m$ phase, our calculations show perfect overlap with the experimental data from Kay et al. \cite{kay_lattice_BaTiO3}.

\begin{figure*}[ht]
    \centering
    \includegraphics[width=0.9\linewidth]{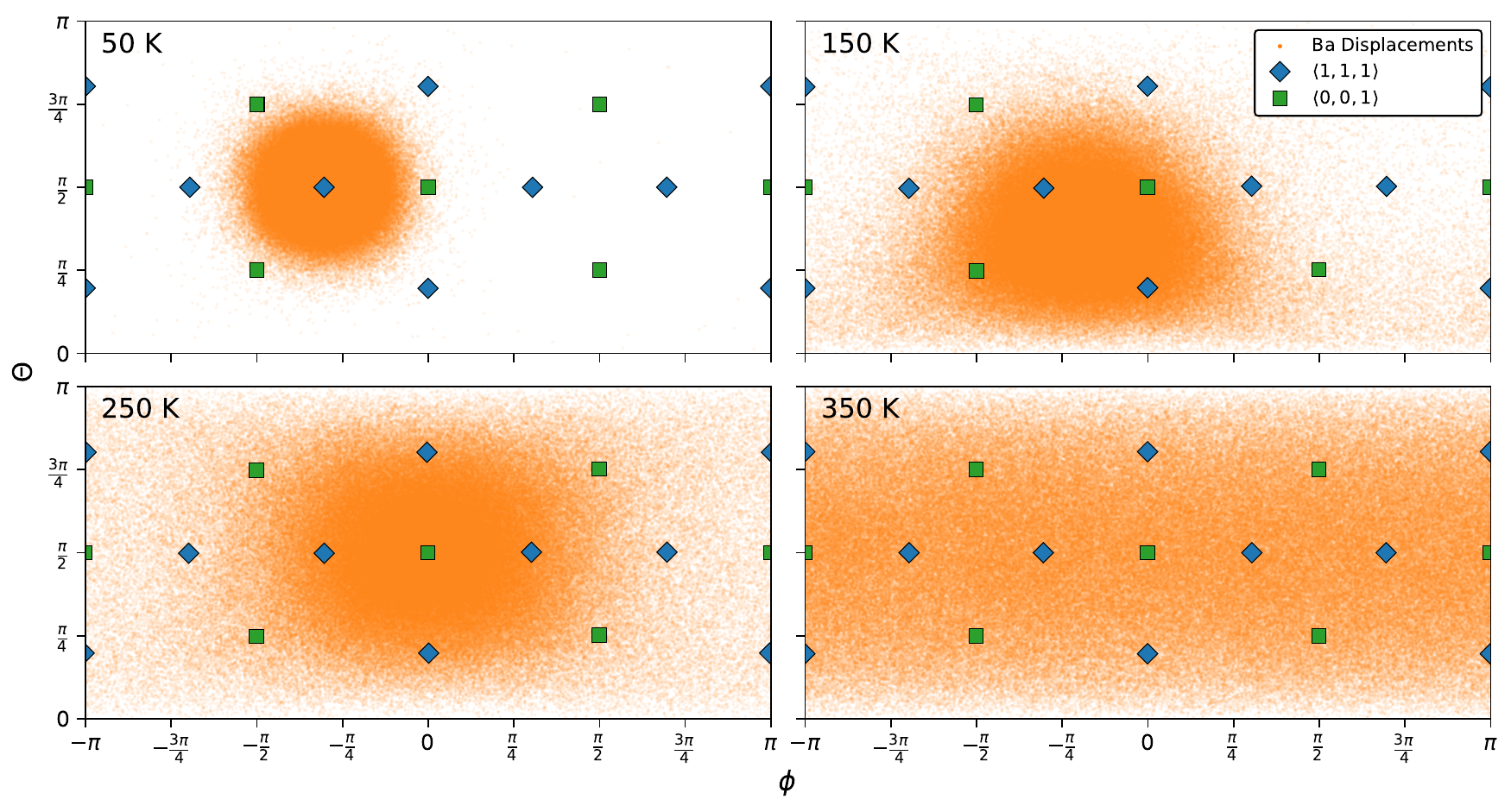}
    \caption{Mercator projections showing the direction of the Ba displacements in \ce{BaTiO3} at $50$ K, $150$ K, $250$ K and $350$ K relative to the $\langle 1,0,0 \rangle$ (green squares) and $\langle 1,1,1 \rangle$ (blue diamonds) directions.}
    \label{fig:mercator_Ba_BaTiO3}
\end{figure*}

Mercator projections showing the direction of Ti displacements at 4 different temperatures are displayed in Figure \ref{main-fig:mercator_Ti_BaTiO3}. SI Figure \ref{fig:mercator_Ba_BaTiO3} displays Mercator projections showing the direction of the Ba displacements at the same temperatures. The directions of the Ba displacements closely follow the Ti displacements, albeit they appear to be a bit more widespread. However, this is likely mainly an artefact of Ba being less displaced than Ti, see SI Figure \ref{fig:abs_all}.

\begin{figure}[ht]
    \centering
    \includegraphics[width=0.70\linewidth]{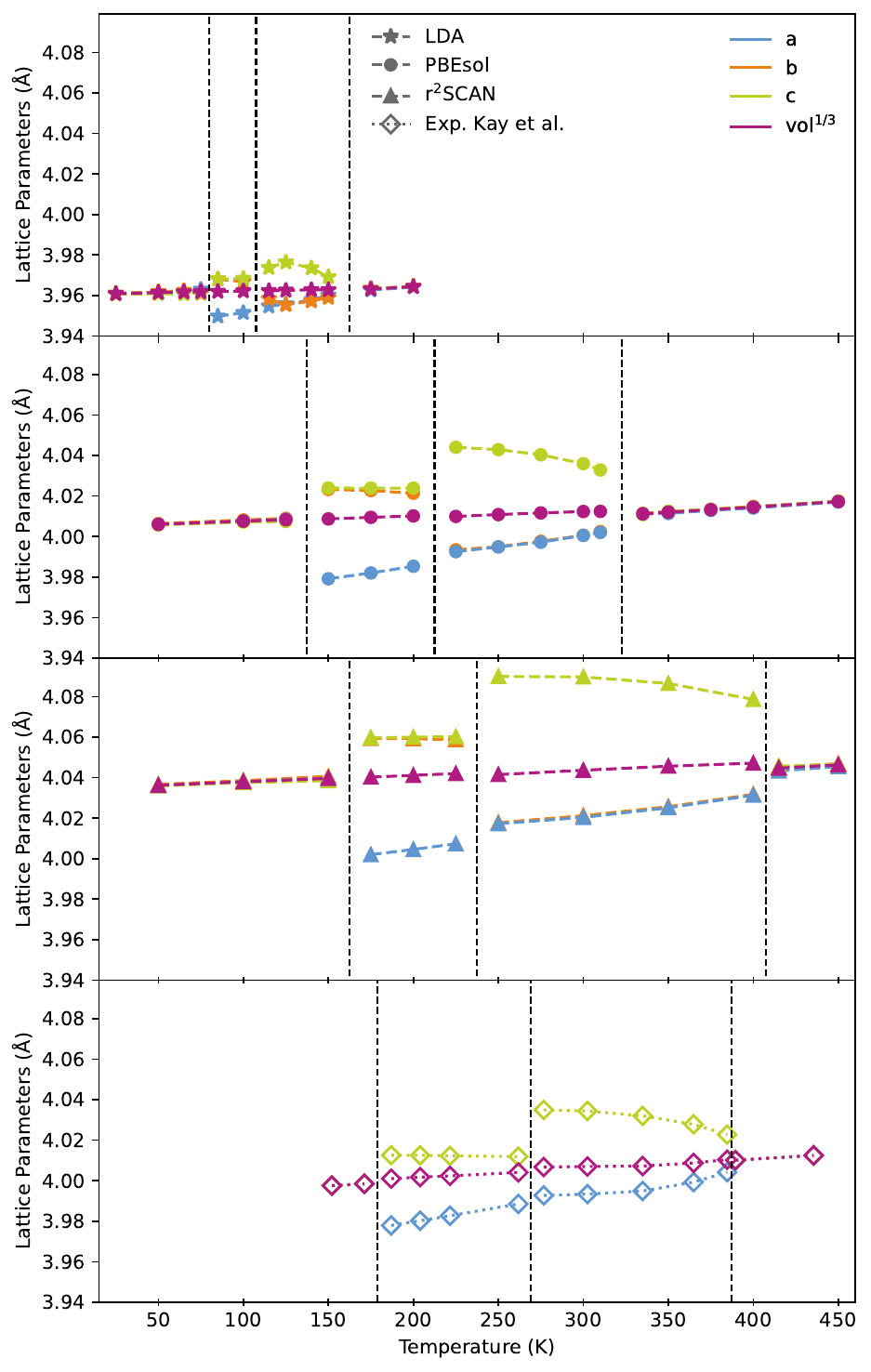}
    \caption{Average lattice parameters as a function of temperature calculated with MLFFs generated using the LDA (first panel), PBEsol (second panel) and r$^2$SCAN (third panel) functional. The fourth panel show experimental data by Kay et al. \cite{kay_lattice_BaTiO3}.}
    \label{fig:BaTiO3_funcs}
\end{figure}

SI Figure \ref{fig:BaTiO3_funcs} shows calculated lattice parameters using MLFFs generated using different functionals for \ce{BaTiO3}. The training of interatomic potentials was carried out using the procedure described in the method section in the main text. The MLFFs predict all the correct phases and phase transitions, albeit with very different lattice parameters and transition temperatures. There is a clear connection between calculated lattice parameter and transition temperatures, where increasing separation between the lattice parameters gives higher transition temperatures. r$^2$SCAN gives transition temperatures closest to the experimental values, but significantly overestimates the lattice parameters. PBEsol results in lattice parameters closest to experimental values. Lattice parameters from LDA are not well described, as the temperature intervals are narrow and close to the transition temperature, lattice parameters tend to fluctuate between the different phases.

\clearpage
\section{P\lowercase{b}T\lowercase{i}O$_{3}$}

\begin{figure}[ht]
    \centering
    \includegraphics[width=0.65\linewidth]{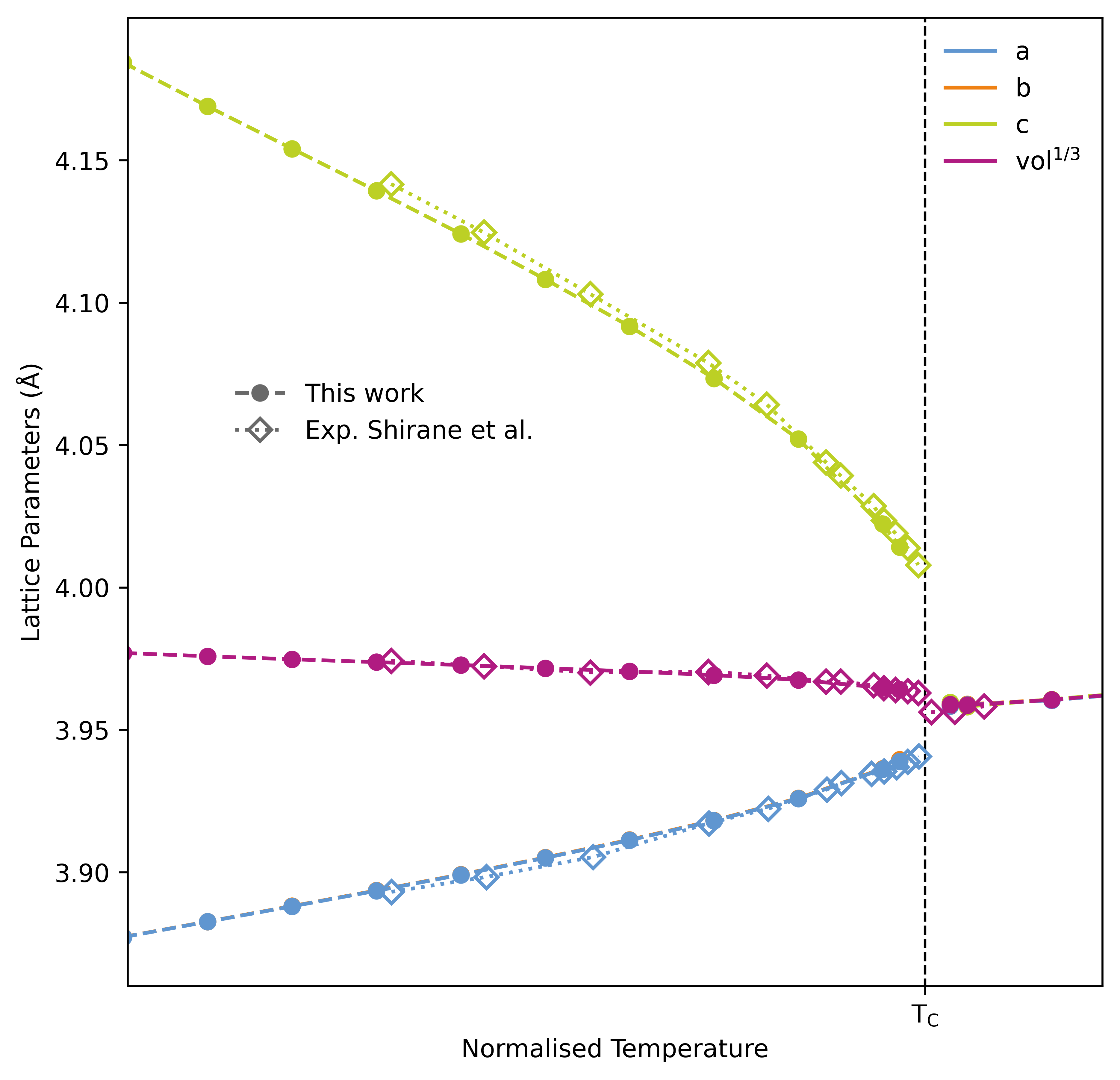}
    \caption{Calculated lattice parameters compared to experimental lattice parameters by Shirane et al. \cite{Shirane_PbTiO3} as a function of temperature normalised to T$_\text{C}$.}
    \label{fig:lattice_PbTiO3_comb}
\end{figure}

Experimental and calculated lattice parameters as a function of temperature normalised to T$_{\text{C}}$, shown in SI Figure \ref{fig:lattice_PbTiO3_comb}, perfectly align both below and above the Curie temperature. 

\begin{figure*}[ht]
    \centering
    \includegraphics[width=0.9\linewidth]{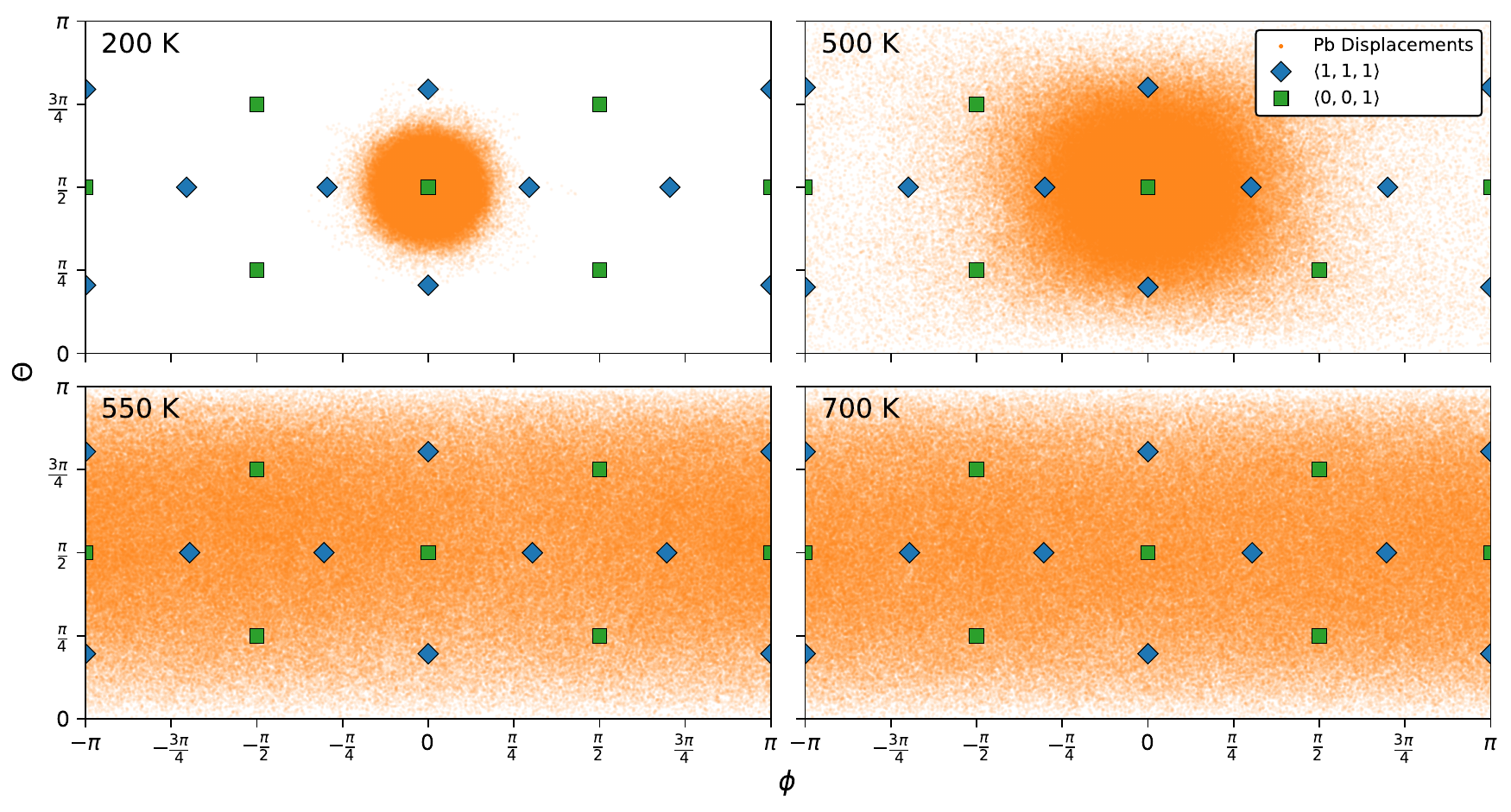}
    \caption{Mercator projections showing the direction of the Pb displacements in \ce{PbTiO3} at $200$ K, $500$ K, $550$ K and $700$ K relative to the $\langle 1,0,0 \rangle$ (green squares) and $\langle 1,1,1 \rangle$ (blue diamonds) directions.}
    \label{fig:mercator_Pb_PbTiO3}
\end{figure*}

\begin{figure*}[ht]
    \centering
    \includegraphics[width=0.9\linewidth]{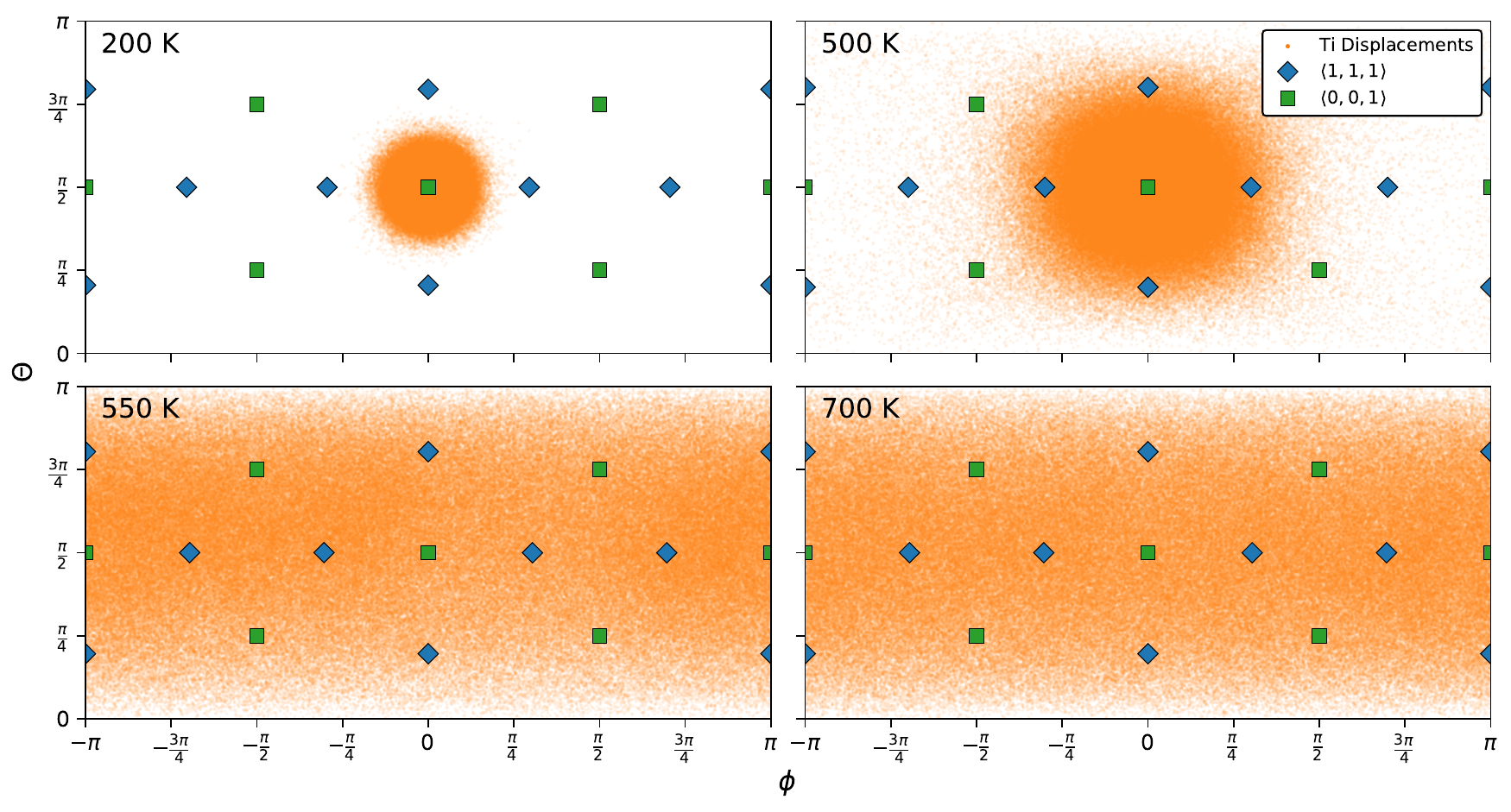}
    \caption{Mercator projections showing the direction of the Ti displacements in \ce{PbTiO3} at $200$ K, $500$ K, $550$ K and $700$ K relative to the $\langle 1,0,0 \rangle$ (green squares) and $\langle 1,1,1 \rangle$ (blue diamonds) directions.}
    \label{fig:mercator_Ti_PbTiO3}
\end{figure*}

Similar to the relation between a- and b-cations in \ce{BaTiO3}, Mercator projections showing the direction of Pb and Ti displacements in \ce{PbTiO3}, SI Figure \ref{fig:mercator_Pb_PbTiO3} and \ref{fig:mercator_Ti_PbTiO3}, show that the Pb displacements follow the Ti displacements, but again appear more widespread. Moreover, in \ce{PbTiO3} the magnitude of the Pb displacements is larger than the Ti displacements and can therefore not explain the difference. However, the Pb displacements may look more widespread as Pb is larger and more polarisable and thus less energy is needed to slightly move it from its equilibrium position.  

\begin{figure}[ht]
    \centering
    \includegraphics[width=0.70\linewidth]{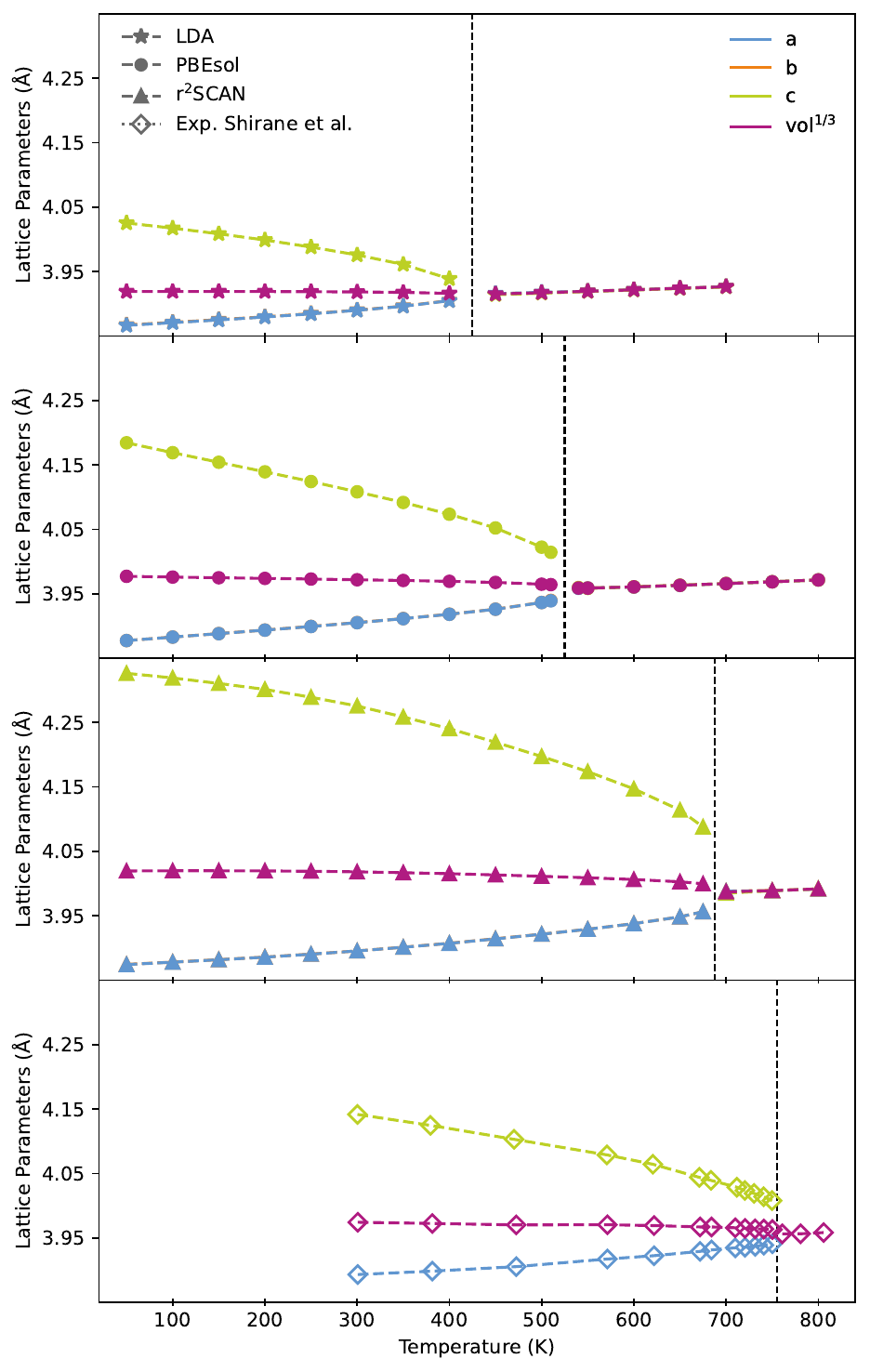}
    \caption{Average lattice parameters as a function of temperature calculated with MLFFs generated using the LDA (first panel), PBEsol (second panel) and r$^2$SCAN (third panel) functional. The fourth panel show experimental data by Shirane et al. \cite{Shirane_PbTiO3}.}
    \label{fig:PbTiO3_funcs}
\end{figure}

Estimated lattice parameters as a function of temperature for \ce{PbTiO3} using interatomic potentials trained using different functionals are displayed in SI Figure \ref{fig:PbTiO3_funcs}. All the MLFFs underestimate the transition temperature. PBEsol results in lattice parameters that are very close to the experimental values, while LDA and r$^2$SCAN severely underestimates and overestimates the lattice parameters, respectively.

\clearpage
\section{L\lowercase{i}N\lowercase{b}O$_{3}$}

\begin{figure}[ht]
    \centering
    \includegraphics[width=0.65\linewidth]{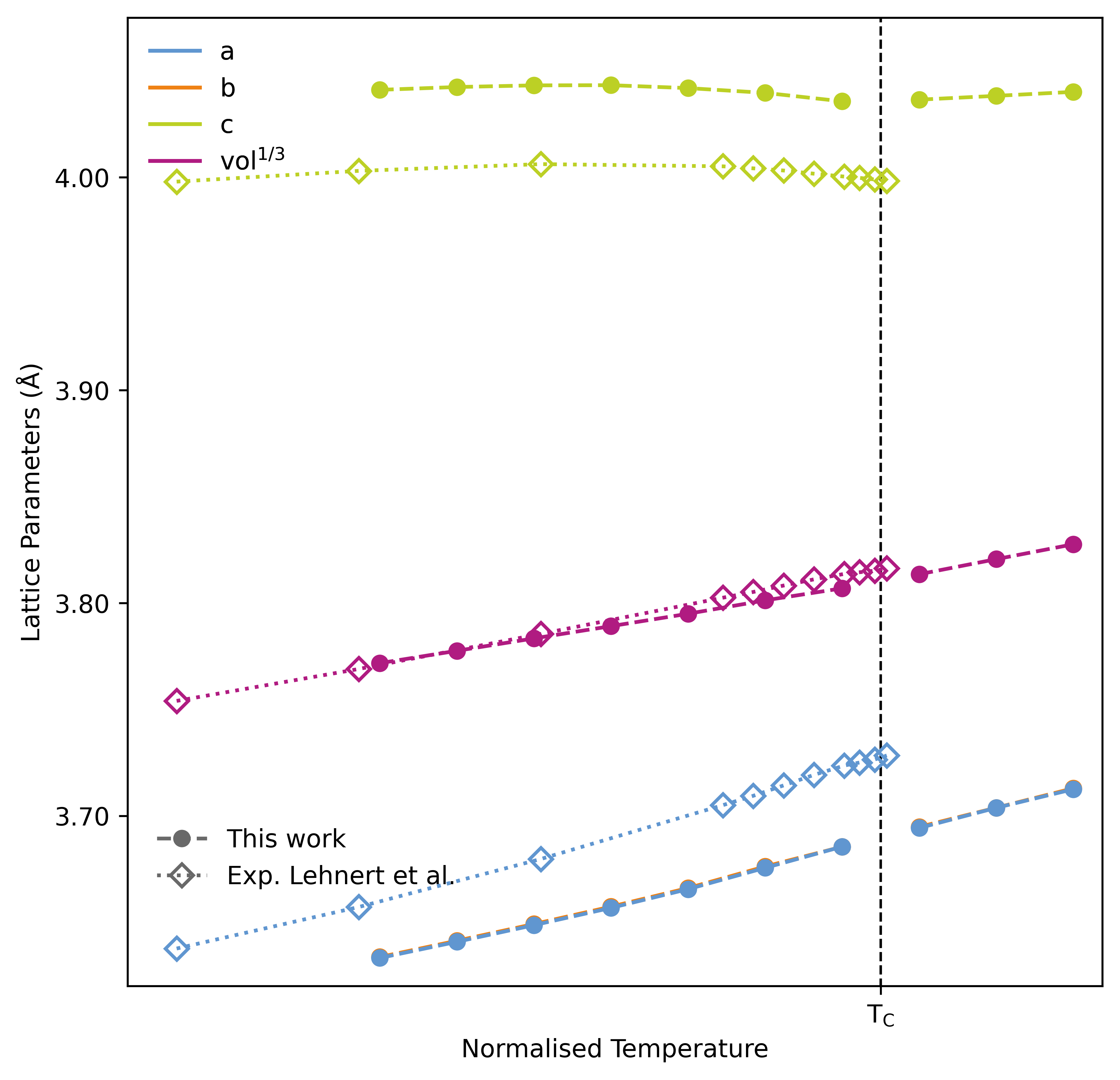}
    \caption{Calculated lattice parameters compared to experimental lattice parameters by Lehnert et al. \cite{LiNbO3_lattice_para} as a function of temperature normalised to T$_\text{C}$. Lattice parameters are converted from the hexagonal conventional unit cell to a pseudocubic cell by dividing a/b and c by $\sqrt{2}$ and $\sqrt{12}$, respectively.}
    \label{fig:lattice_LiNbO3_comb}
\end{figure}

Calculated and experimental lattice parameters as a function of temperature normalised to T$_{\text{C}}$, displayed in SI Figure \ref{fig:lattice_LiNbO3_comb}, show a similar evolution of the structure with temperature. The calculated c-parameters are systematically larger than the experimental values, while the a-parameters are shorter. However, there is an almost perfect overlap between the calculated and experimental volumes.

\begin{figure*}[ht]
    \centering
    \includegraphics[width=0.9\linewidth]{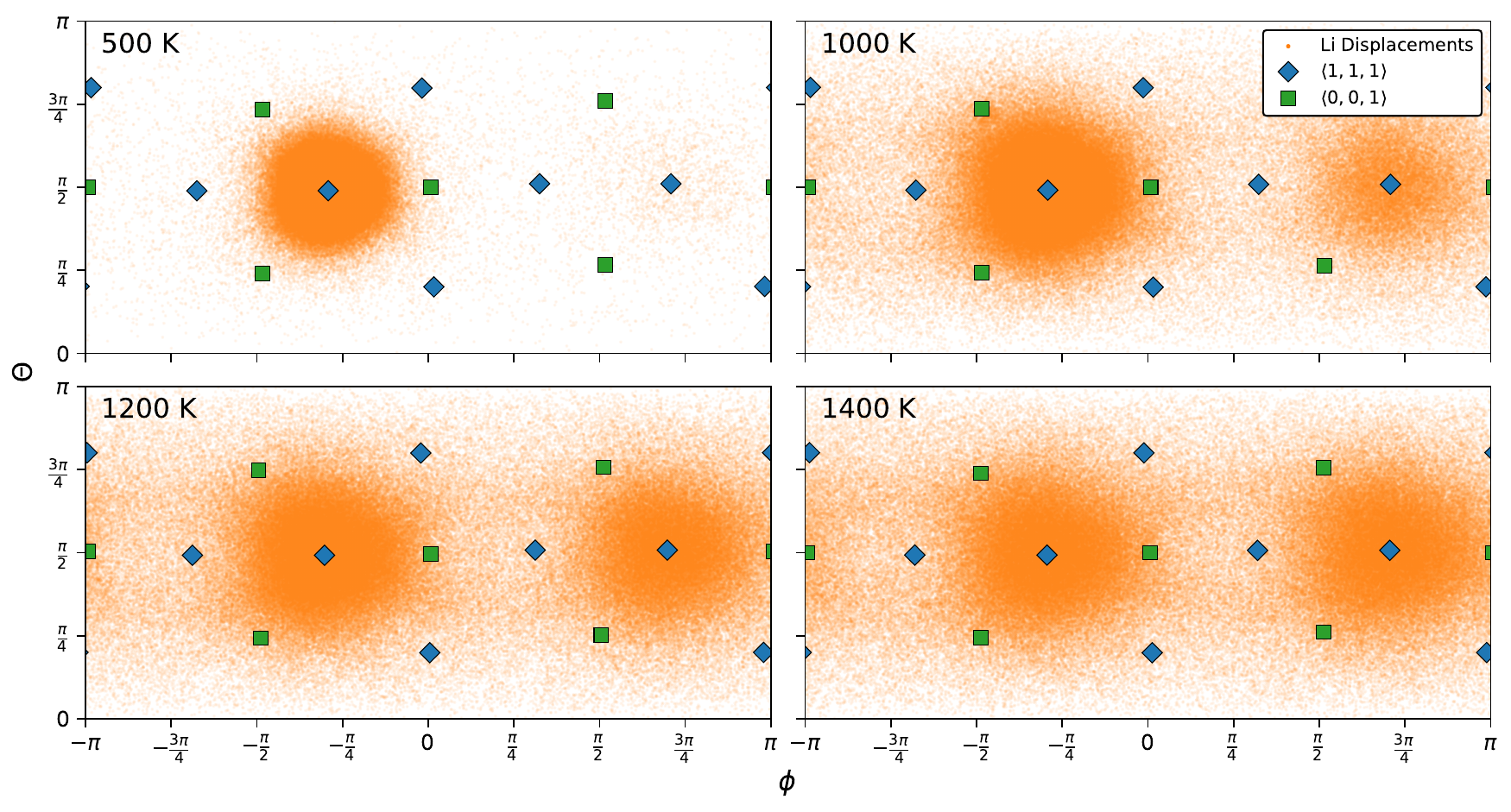}
    \caption{Mercator projections showing the direction of the Li displacements in \ce{LiNbO3} at $500$ K, $1000$ K, $1200$ K and $1400$ K relative to the $\langle 1,0,0 \rangle$ (green squares) and $\langle 1,1,1 \rangle$ (blue diamonds) directions.}
    \label{fig:mercator_Li_LiNbO3}
\end{figure*}

\begin{figure*}[ht]
    \centering
    \includegraphics[width=0.9\linewidth]{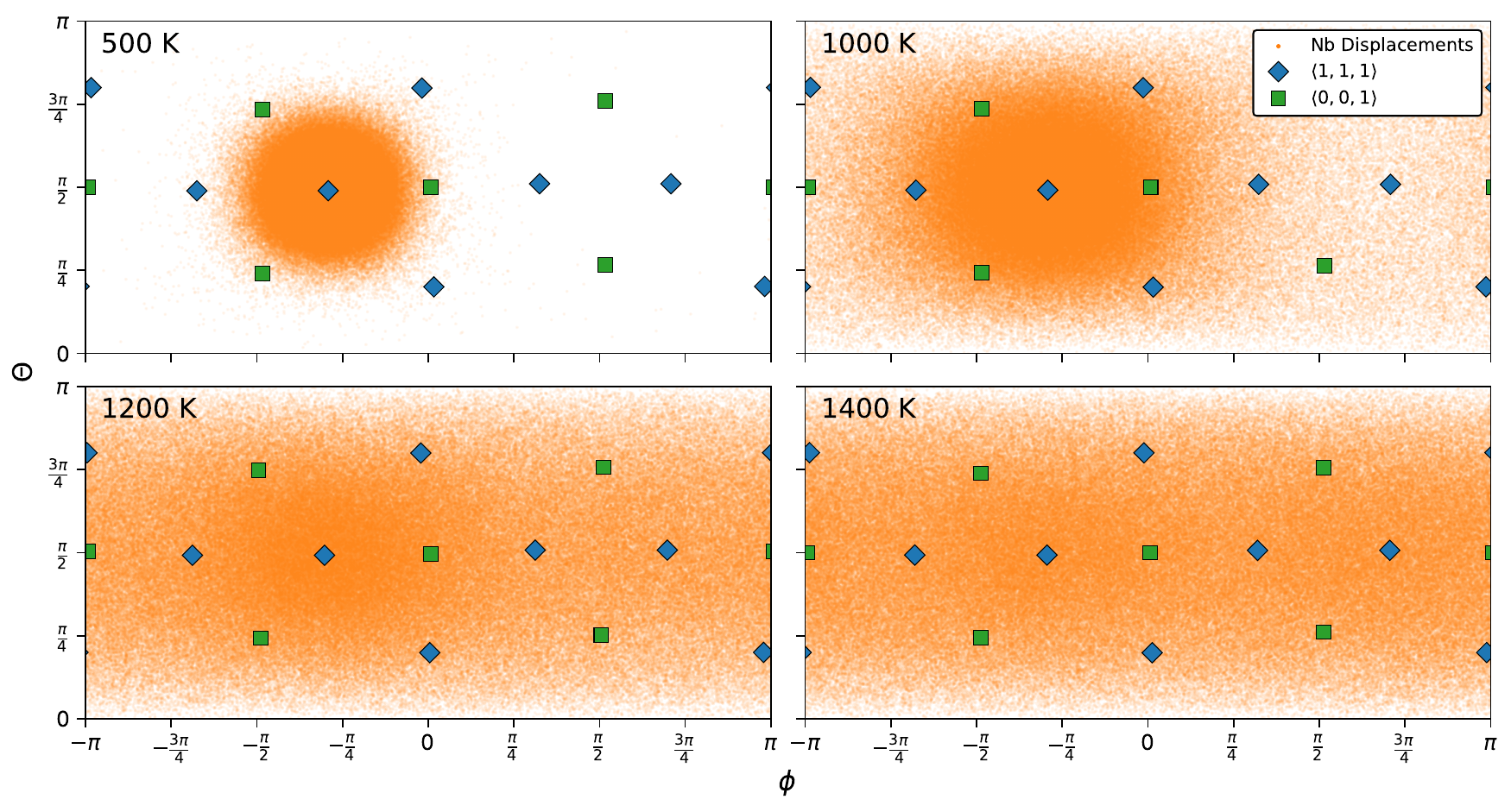}
    \caption{Mercator projections showing the direction of the Nb displacements in \ce{LiNbO3} at $500$ K, $1000$ K, $1200$ K and $1400$ K relative to the $\langle 1,0,0 \rangle$ (green squares) and $\langle 1,1,1 \rangle$ (blue diamonds) directions.}
    \label{fig:mercator_Nb_LiNbO3}
\end{figure*}

The Mercator projections in SI Figure \ref{fig:mercator_Li_LiNbO3} clearly show the order-disorder transition of the Li displacements. At low temperatures, Li is displayed along one of the $\langle1\times1\times1\rangle$ directions, while at higher temperatures the displacements are split between two opposite $\langle1\times1\times1\rangle$ directions. Mercator projections for Nb, Si Figure \ref{fig:mercator_Nb_LiNbO3}, show Nb following a displacive transition.

\begin{figure}[ht]
    \centering
    \includegraphics[width=0.70\linewidth]{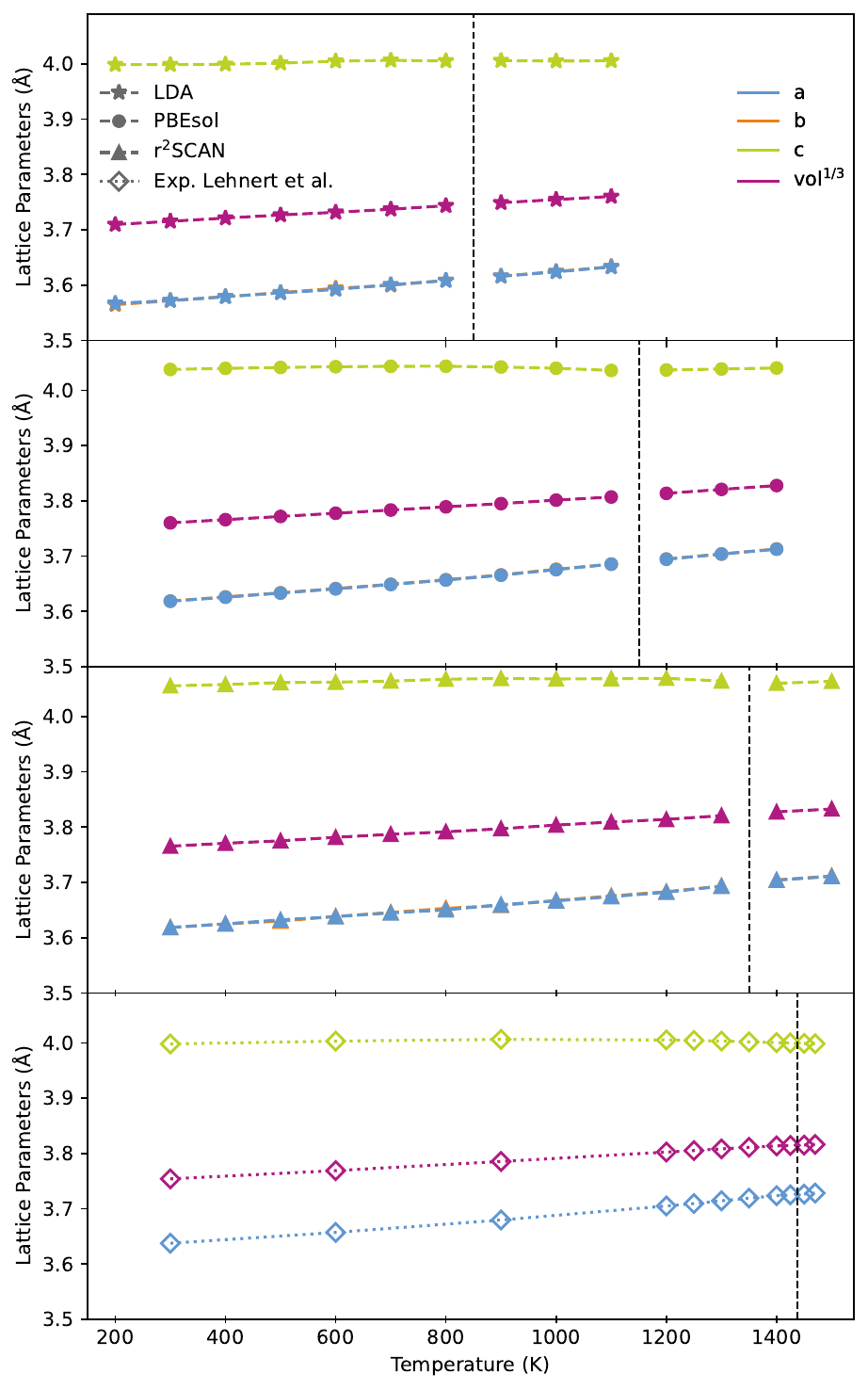}
    \caption{Average lattice parameters as a function of temperature calculated with MLFFs generated using the LDA (first panel), PBEsol (second panel) and r$^2$SCAN (third panel) functional. The fourth panel show experimental data by Lehnert et al. \cite{LiNbO3_lattice_para}.}
    \label{fig:LiNbO3_funcs}
\end{figure}

SI Figure \ref{fig:LiNbO3_funcs} shows calculated lattice parameters using MLFFs generated using different functionals for \ce{LiNbO3}. The different interatomic potentials result in the correct phases, but the different functionals all underestimate the transition temperature.

\clearpage
\section{B\lowercase{i}F\lowercase{e}O$_{3}$}

\begin{figure}[ht]
    \centering
    \includegraphics[width=0.65\linewidth]{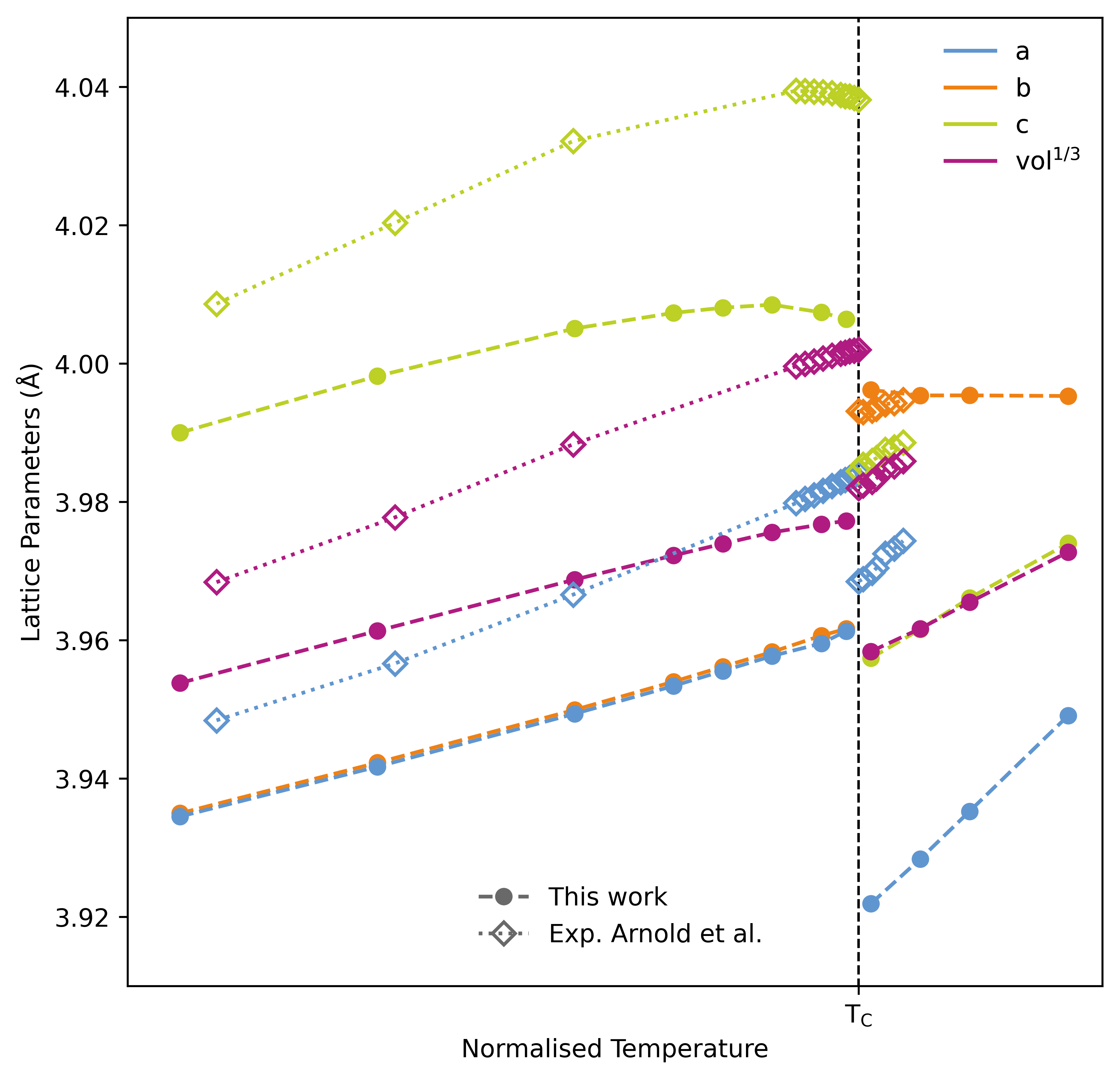}
    \caption{Calculated lattice parameters compared to experimental lattice parameters by Arnold et al. \cite{arnold_BiFeO3} as a function of temperature normalised to T$_\text{C}$. Similar to \ce{LiNbO3}, in the rhombohedral phase, lattice parameters are converted from the hexagonal conventional unit cell to a pseudocubic cell by dividing a/b and c by $\sqrt{2}$ and $\sqrt{12}$, respectively. Lattice parameters from the orthorhombic structure are converted by dividing a/b and c by $\sqrt{2}$ and $2$, respectively.}
    \label{fig:lattice_BiFeO3_comb}
\end{figure}

SI Figure \ref{fig:lattice_BiFeO3_comb} shows large discrepancies between calculated and experimental lattice parameters of \ce{BiFeO3} for both the ferroelectric and paraelectric phases. However, the trends show a relatively good match. Additionally, calculated lattice parameters closely follow lattice parameters expected using standard DFT when considering thermal expansion. DFT optimised lattice parameters for the $Pbnm$ structure are shown in Table \ref{tab:BiFeO3_lattice_Pbnm} and are relatively similar to the lattice parameters calculated using MLFFs at elevated temperatures.

\begin{figure*}[ht]
    \centering
    \includegraphics[width=0.9\linewidth]{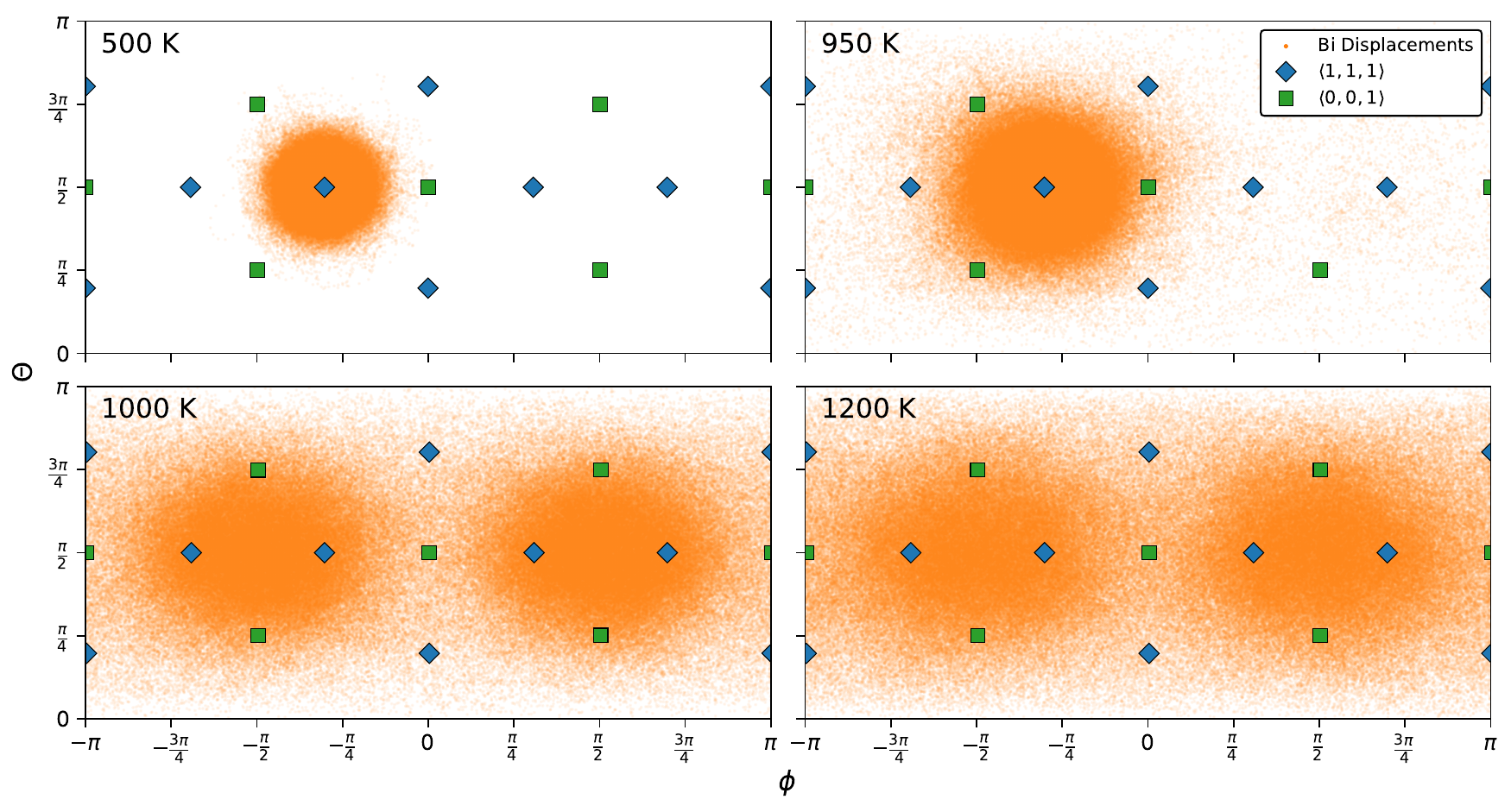}
    \caption{Mercators projection showing the direction of the Bi displacements in \ce{BiFeO3} at $500$ K, $950$ K, $1000$ K and $1200$ K relative to the $\langle 1,0,0 \rangle$ (green squares) and $\langle 1,1,1 \rangle$ (blue diamonds) directions.}
    \label{fig:mercator_Bi_BiFeO3}
\end{figure*}

\begin{figure*}[ht]
    \centering
    \includegraphics[width=0.9\linewidth]{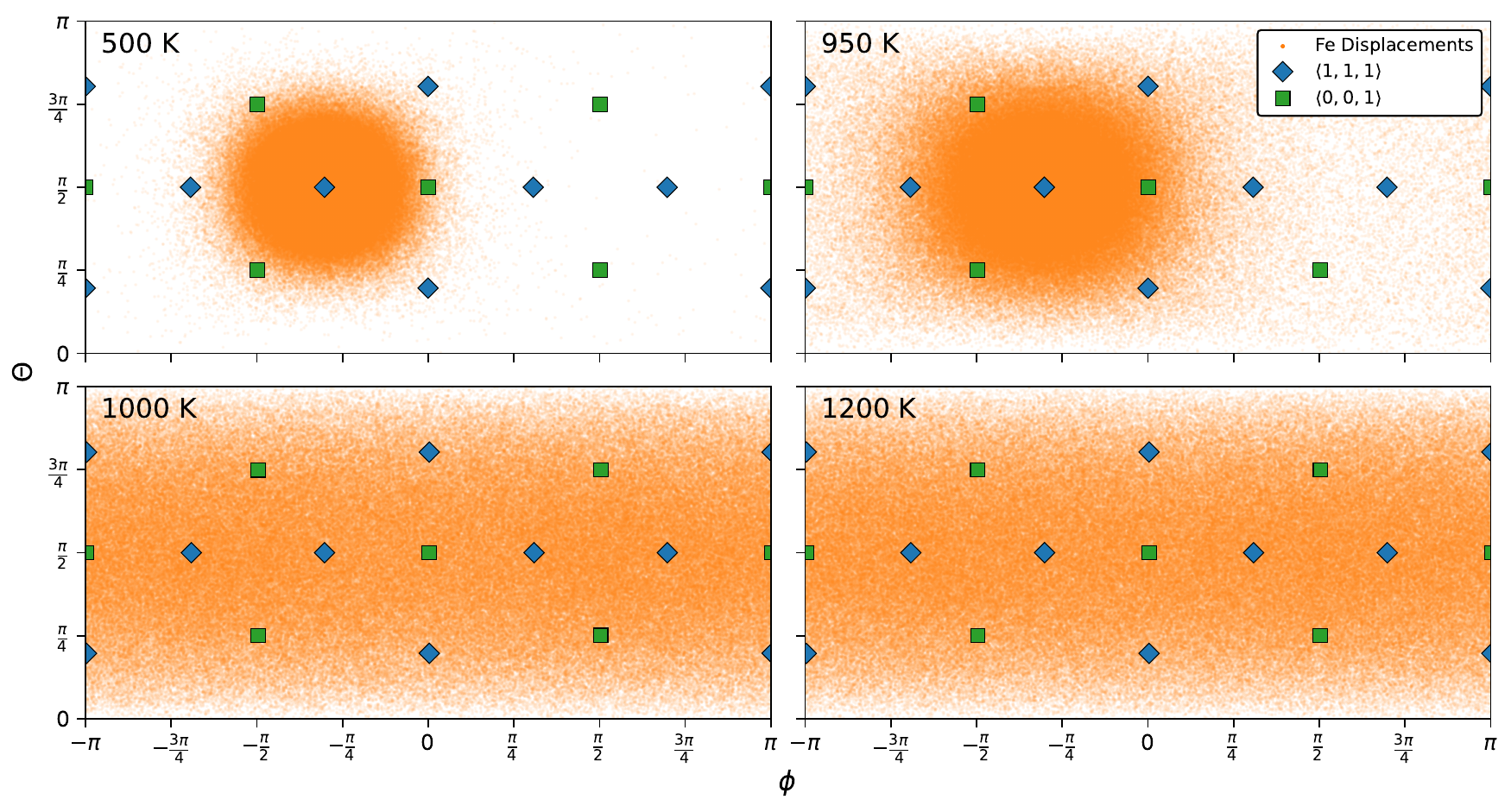}
    \caption{Mercator projections showing the direction of the Fe displacements \ce{BiFeO3} at $500$ K, $950$ K, $1000$ K and $1200$ K relative to the $\langle 1,0,0 \rangle$ (green squares) and $\langle 1,1,1 \rangle$ (blue diamonds) directions.}
    \label{fig:mercator_Fe_BiFeO3}
\end{figure*}

The Mercator projections for Bi in \ce{BiFeO3}, shown in SI Figure \ref{fig:mercator_Bi_BiFeO3}, appear very similar to Li in \ce{LiNbO3}. However, at elevated temperatures, Bi displacements are not disordered between two directions, but instead ordered alternatingly. On the other hand, Fe shows a very abrupt change from being displaced in the ferroelectric phase to not being displaced above the T$_{\text{C}}$, see SI Figure \ref{fig:mercator_Fe_BiFeO3}.

\begin{table}[ht]
    \centering
    \caption{DFT optimised lattice parameters for $Pbnm$ \ce{BiFeO3} using the PBEsol functional.}
    \begin{tabular}{cc}
        \hline
        a & 5.415 \\
        b & 5.641 \\
        c & 7.793 \\
        $\alpha$, $\beta$, $\gamma$ & $90^{\circ}$ \\
        \hline
    \end{tabular}
    
    \label{tab:BiFeO3_lattice_Pbnm}
\end{table}

\begin{figure}[ht]
    \centering
    \includegraphics[width=0.7\linewidth]{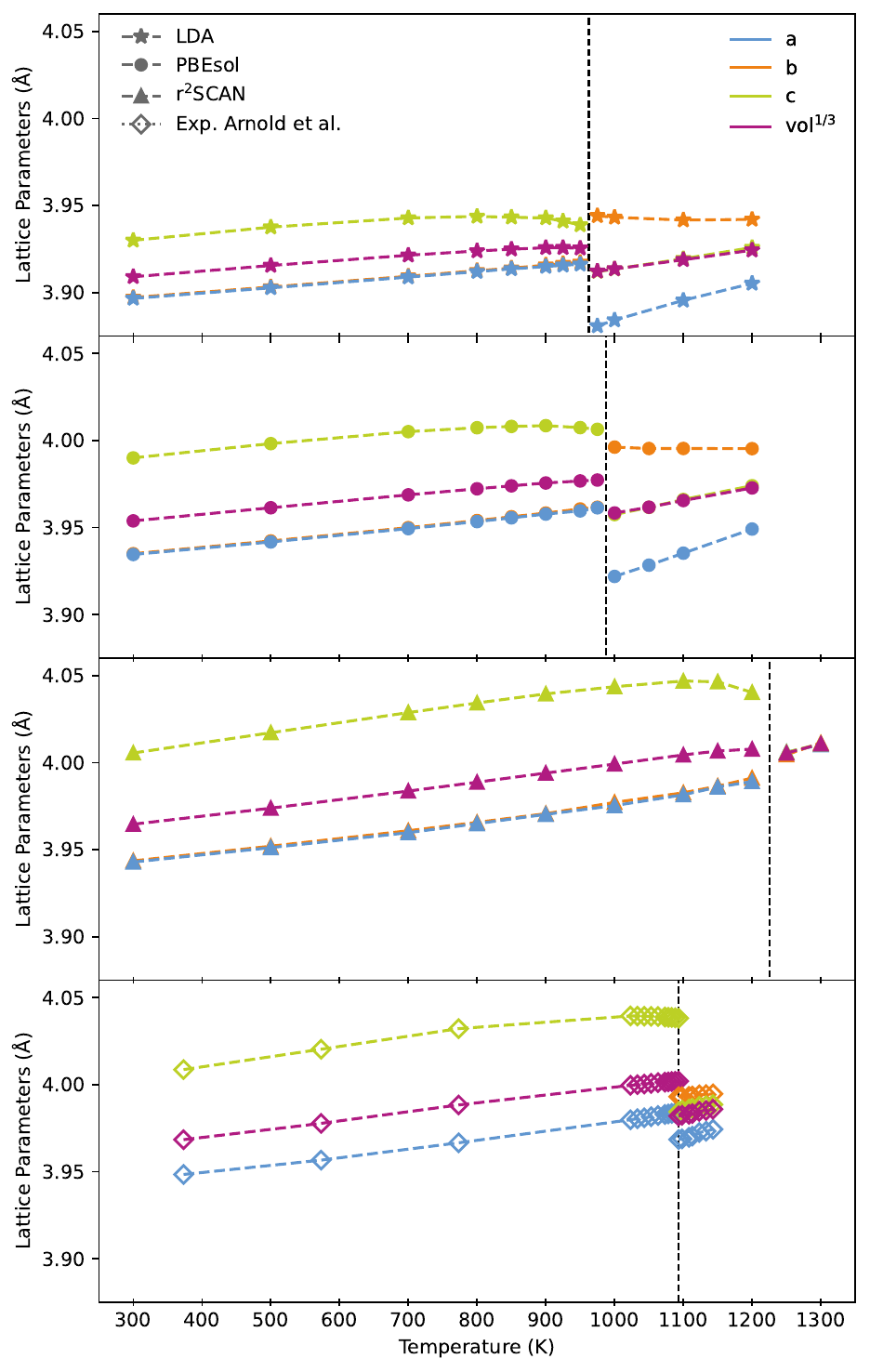}
    \caption{Average lattice parameters as a function of temperature calculated with MLFFs generated using the LDA (first panel), PBEsol (second panel) and r$^2$SCAN (third panel) functional. The fourth panel show experimental data by Arnold et al. \cite{arnold_BiFeO3}.}
    \label{fig:BiFeO3_funcs}
\end{figure}

Lattice parameters calculated using MLFFs generated with LDA, PBEsol and r$^2$SCAN are shown in SI Figure \ref{fig:BiFeO3_funcs} together with experimental lattice parameters by Arnold et al. \cite{arnold_BiFeO3}. LDA and PBEsol underestimates the lattice parameters, but show a transition from $R3c$ to $Pbnm$ at a temperature relatively close to the experimental Curie temperature. On the other hand, r$^2$SCAN shows a transition from $R3c$ to $Pm\Bar{3}m$ at a temperature far above the experimental T$_C$.

\clearpage

\bibliography{references}